\documentclass{article}
\usepackage{arxiv}

\usepackage[utf8]{inputenc} \usepackage[T1]{fontenc}    \usepackage{hyperref}       \usepackage{url}            \usepackage{booktabs}       \usepackage{amsfonts}       \usepackage{nicefrac}       \usepackage{microtype}      \usepackage{lipsum}
\usepackage{graphicx}
\usepackage{natbib}
\usepackage{caption} 
\usepackage{xspace}
\usepackage{amsmath}
\usepackage{cleveref}
\usepackage{subcaption}
\usepackage{soul}
\usepackage{xcolor}
\usepackage{enumitem}
\usepackage{parskip}
\usepackage{lmodern}

\usepackage{booktabs}
\usepackage{xspace}

\graphicspath{ {./AAAI_2027Template/AnonymousSubmission/}{./AAAI_2027Template/Supplementary Materials/} }
\definecolor{dodgerblue}{HTML}{1E90FF}
\definecolor{linkblue}{HTML}{0000FF}
\definecolor{white}{HTML}{FFFFFF}

\newcommand{\Emails}{\texttt{PhishFuzzer}\xspace}
\newcommand{\Cub}{\texttt{CUB}\xspace}
\newcommand{\SM}{SM\xspace}

\crefname{figure}{Fig.}{Figs.}
\newcommand{\separator}{\noindent\texttt{-------}}

\definecolor{blueish}{HTML}{4287f5}
\definecolor{anotherpinkish}{HTML}{8a2456}
\definecolor{black}{HTML}{0d0d0d}
\hypersetup{
    colorlinks=true,
    linkcolor=black,
    citecolor=anotherpinkish,
    urlcolor=anotherpinkish,
}

\title{Are Concept Bottleneck Models Effective as Decision-Support Systems?}
\author{
 Alessandro Bogani \footnotemark[1] \\
  DISI, University of Trento, Italy\\
  \texttt{alessandro.bogani@unitn.it} \\
   \And
 Nicola Debole \thanks{Shared first author.}\\
  DISI, University of Trento, Italy\\
  \texttt{nicola.debole@unitn.it} \\
   \And
 Emanuele Marconato \\
  DISI, University of Trento, Italy\\
  \texttt{emanuele.marconato@unitn.it} \\
  \And
 Andrea Pugnana \\
  DISI, University of Trento, Italy\\
  \texttt{andrea.pugnana@unitn.it} \\
  \And
 Katya Tentori \\
CIMeC, University of Trento, Italy\\
  \texttt{katya.tentori@unitn.it} \\
  \And
 Andrea Passerini\hspace{0.2em}\\
  DISI, University of Trento, Italy\\
    \texttt{andrea.passerini@unitn.it} \\
}

\begin{document}
\maketitle
\begin{abstract}
Concept Bottleneck Models (CBMs) are interpretable-by-design neural networks that detect human-understandable concepts from the input and use them to generate predictions. By allowing users to inspect the concepts underlying a prediction and explore how predictions change under alternative concept configurations, CBMs have emerged as one of the most prominent approaches to supporting human–AI collaboration. 
However, user studies investigating their actual effectiveness as decision-support systems remain limited. We present two large-scale user studies (\textit{N\textsubscript{participants}} = 705, \textit{N\textsubscript{observations}} = 6,959) evaluating how concept-based explanations and user interventions on the model’s concepts affect the performance of the human-AI team in two distinct binary classification tasks.
Our results show that CBMs, and particularly their interactive component, can improve human–AI team accuracy relative to both unaided human performance and performance with non-interpretable AI support. However, these benefits emerge only under certain conditions: classification tasks perceived as difficult, easily identifiable concepts, and active interaction with the model. We also discuss how inaccurate concept detection may undermine users’ trust in the model. Overall, this work provides practical guidance for the deployment of CBMs as effective decision-support tools. Code and data are available at \href{https://github.com/debryu/user-study-CBMs}{https://github.com/debryu/user-study-CBMs}.
\end{abstract}

\addtocontents{toc}{\protect\setcounter{tocdepth}{0}}
\section{Introduction}
\addtocontents{toc}{\protect\setcounter{tocdepth}{2}}

The lack of interpretability in Artificial Intelligence (AI) systems is widely recognized as a major barrier to their effective use 
\citep{DBLP:journals/cogcom/HassijaCMSGHSSMH24}. 
Concept Bottleneck Models (CBMs)~\citep{DBLP:conf/icml/KohNTMPKL20} offer an interpretable-by-design approach to mitigate this issue by basing predictions on human-understandable concepts, and have thus received increasing attention in recent years~\citep{DBLP:journals/tmlr/KnabSBKSSSS26}.

The appeal of CBMs is twofold: detected concepts can serve as explanations for the model's predictions and, most importantly, they allow users to interactively examine how predictions would change under different concept configurations~\citep {DBLP:conf/iclr/DominiciBGGML25}, a component generally regarded as enhancing explanatory approaches \citep{DBLP:journals/frai/TesoASD23}. Together, these properties are assumed to foster appropriate reliance on the model by helping users identify cases in which it has incorrectly detected or failed to detect the presence of a concept, and by enabling them to assess the robustness of its predictions to modifications to the concept set.

However, empirical evidence supporting the effectiveness of concept-based explanatory approaches, including CBMs, remains very limited~\citep{poeta2025concept}.
In particular, it is still unclear whether CBMs are effective as decision-support systems and how the ability to inspect or intervene on detected concepts affects users' reliance on the system's predictions, confidence in their own judgments, and trust in the model.

In this work, we address this gap through two user studies evaluating CBMs as decision-support systems. To better identify the conditions under which they are most effective, we consider classification tasks across two contexts with complementary properties: one in which users are familiar with the objects to be classified, but identifying the relevant concepts involves a degree of subjectivity (i.e., classifying emails as legitimate or fraudulent based on emotional and goal-oriented cues), and another in which users are less familiar with the objects to be classified, but concept identification is more objectively grounded (i.e., classifying birds as one of two sparrow species based on visual features). 

Depending on the experimental condition, participants complete these tasks with different levels of support from the CBM, ranging from no assistance at all to progressively richer forms of support: the predicted label; the predicted label together with the detected concepts; and the predicted label together with the detected concepts, as well as the possibility for users to interact with them. Our results show that CBMs, and in particular their interactive component, can improve human–AI team performance beyond both unaided human performance and performance with non-interpretable AI support, but only under specific conditions: classification tasks perceived as difficult, easily identifiable concepts, and active interaction with the model. Finally, our findings also suggest that inaccurate concept detection may, to some extent, undermine users' trust in the model.

\begin{figure*}[t]
\centering
\includegraphics[width=16cm]{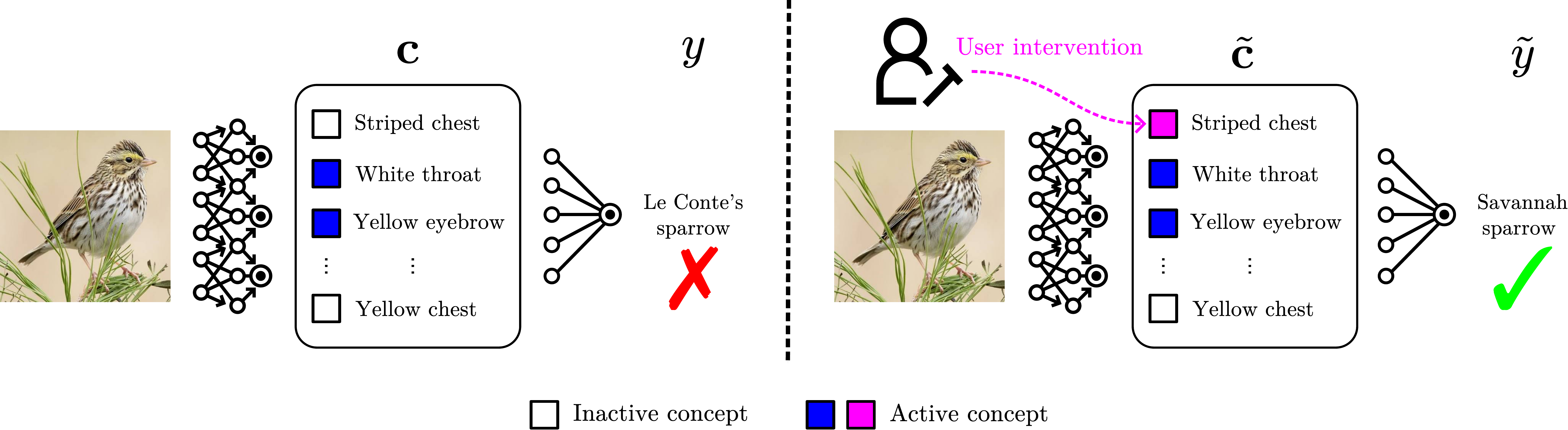}
\caption{
\textbf{Overview of the Concept Bottleneck Model and its interactive component}. 
(Left) A concept encoder determines the presence of human-interpretable concepts $\mathbf c$ in the input image, which are passed to a task predictor to generate a classification label $y$. Here, because \textit{striped chest} is erroneously not detected in the bird image, the final result is incorrect.   
(Right)
Users can intervene on the detected concepts by changing the predicted values to a new $\tilde{\mathbf c}$, here setting \textit{striped chest} to be active.
After this step, the prediction is updated according to the new concept values, leading to a correct task prediction $\tilde y$.
}
\label{fig:overview}
\end{figure*}

\paragraph{Contributions.}
Our main contributions are as follows:
\begin{itemize}
    \item[$(i)$] We introduce a new experimental paradigm that can be applied across a variety of datasets to evaluate whether, and to what extent, CBMs improve human–AI collaboration in terms of overall accuracy, users’ confidence in their decisions, and their trust in the AI system.
    
    \item[$(ii)$] We conduct two large-scale user studies involving 705 participants and two datasets with distinct characteristics. Based on the results of these studies, we provide practical guidance on the conditions under which CBMs are most likely to improve human–AI collaboration.
\end{itemize}

\addtocontents{toc}{\protect\setcounter{tocdepth}{0}}
\section{Background}
\addtocontents{toc}{\protect\setcounter{tocdepth}{2}}

\paragraph{Concept Bottleneck Models.}
CBMs~\citep{DBLP:conf/icml/KohNTMPKL20} are  interpretable-by-design models that provide label predictions by composing 
(i) a \textit{concept encoder} $g: \mathcal{X} \to [0,1]^{n_c}$ that maps inputs $\mathbf{x} \in \mathcal{X} \subseteq \mathbb{R}^d$ (e.g., an image) to a set of $n_c$ binary concept activations $\mathbf{c} \in \mathcal{C} = [0,1]^{n_c}$ (e.g., the probability that the image contains a ``\texttt{red}'' object), 
and (ii) a \textit{task predictor} $f: \mathcal{C} \to \mathcal{Y}$ that maps, \textit{in a linear manner}, concept activations to the label prediction $y \in \mathcal{Y} = \{1, \ldots, n_y \}$ (e.g., whether the input is an ``\texttt{apple}'' or a ``\texttt{pear}''). 

Most CBMs are trained by leveraging  dense supervision on both active and inactive concepts and label predictions for the input. 
The concept encoder $g$ and the task predictor $f$ can be trained following different approaches:  (a) \textit{independently}, i.e., $g$ and $f$ are trained separately; (b) \textit{sequentially}, i.e., one first trains $g$ and then uses its output to train $f$; or (c) \textit{jointly}, i.e., $g$ and $f$ are trained together at the same time.

At inference time, CBMs allow users to intervene on the detected concepts thanks to their modular structure. For an input $x\in \mathcal{X}$, the label prediction is obtained as ${y} = f(g(x))$ (see \cref{fig:overview}, left). Users can then inspect the concept activations $g(x)$ and override any subset of them with values of their own, yielding a modified concept vector $\tilde{\mathbf c}$. The label prediction is then updated as $\tilde{y} = f(\tilde{\mathbf c})$ (see \cref{fig:overview}, right).

\addtocontents{toc}{\protect\setcounter{tocdepth}{0}}
\section{Present Work}
\addtocontents{toc}{\protect\setcounter{tocdepth}{2}}

We evaluate the effectiveness of CBMs in supporting participants in binary classification tasks through two large-scale studies. In both studies, we ask participants to provide a binary classification of several items, either with or without the support of a CBM. In Study 1, the task consists of classifying emails as fraudulent or legitimate, whereas in Study 2, it involves classifying images of birds as belonging to one of two sparrow species. Because the two studies share a similar experimental design, we describe their methods in a single section. 
The research protocol for both studies was submitted to the Research Ethics Committee of [institution omitted to preserve anonymity], which determined that the studies posed no risk to participants’ well-being or rights and therefore did not require full ethical review ([protocol number omitted to preserve anonymity]). All materials required to reproduce our experiments or to reuse our experimental paradigms, together with the associated data, are provided in the accompanying Code and Data Supplement.

\addtocontents{toc}{\protect\setcounter{tocdepth}{0}}
\subsection{Datasets and CBMs specifics}
\addtocontents{toc}{\protect\setcounter{tocdepth}{2}}

\paragraph{Datasets.}
We use two datasets to train the CBMs, each providing ground-truth annotations for both the class labels and the human-interpretable features used as concepts in the CBMs' bottleneck. As detailed below, we select six features from each dataset (three predictive of one class and three predictive of the other) to serve as bottleneck concepts. Restricting the bottleneck to six concepts is intended to reduce participants’ cognitive load, in line with prior work suggesting that CBMs relying on a large number of concepts may be perceived as impractical~\citep{DBLP:conf/cvpr/RamaswamyKFR23}.

The first dataset, employed in Study 1, is \Emails \citep{toth2025phishfuzzer}, which comprises a corpus of human-written emails together with LLM-generated rephrasings of them. Each email is annotated both for its class ("valid", "phishing", or "spam") and for a range of content features, including attempts to elicit specific emotional states (e.g., "fear" or "urgency") or to prompt particular actions from the recipient (e.g., "open attachment" or "reply").
For our purposes, we consider only human-written emails in English belonging to either the "valid" or "phishing" classes, relabeled as "legitimate" and "fraudulent", respectively (see Supplementary Materials, henceforth \SM, for the full list of email selection criteria).
Regarding the features annotated in the dataset, the ones considered for the concepts in the bottleneck of the CBM are "problem alert", "time pressure", "attachment interaction", "update notification", "operational tone", and "reply request" (in both Studies 1 and 2, some dataset feature names are relabeled when used as CBM concept names to reduce potential ambiguity for participants; see \SM for the selection criteria and descriptions of the concepts used in each study).

The second dataset, used in Study 2, is \Cub \citep{WahCUB_200_2011}, one of the most widely used datasets in the CBM literature. It comprises bird images annotated with both species labels and a set of visual attributes. We select two sparrow species ("Le Conte's" and "Savannah") whose visual features (i.e., the concepts) made the classification task manageable, while still preserving a meaningful level of difficulty and leaving room for AI support to improve performance. Among the features annotated in \Cub, the ones considered for the bottleneck of the CBM are: "warm-colored eyebrow", "warm-colored chest", "plain sides", "crested head", "white throat", and "striped chest". 

\paragraph{CBMs specifics.}
As depicted in \cref{fig:overview}, our CBM implementation features a deep \textit{concept encoder} and a shallow linear \textit{task predictor}, which are trained independently. 
The \textit{concept encoder} is composed of a frozen neural encoder, which maps raw inputs to latent embeddings, followed by $n_c$ (one for each concept) binary Support Vector Machine (SVM) classifiers~\citep{DBLP:journals/ml/CortesV95}.
The output of the SVMs returns the predicted concepts of the CBM (refer to \SM for implementation details).
The \textit{task predictor} is a simple logistic regression model that outputs (binary) class probabilities. 

The concept extractor is trained using 6,384 and 28,470 ground-truth concept values for \Emails and CUB respectively. The task extractor is trained with 366 phishing and 185 valid emails for \Emails; 25 Le Conte's Sparrow and 26 Savannah Sparrow for \Cub.
Complete implementation details are provided in the \SM.

The CBM trained on the \Emails achieves $92.3\%$ test accuracy, while the CBM trained on the CUB dataset achieves $81.4\%$ test accuracy.

\addtocontents{toc}{\protect\setcounter{tocdepth}{0}}
\subsection{Methodology}
\addtocontents{toc}{\protect\setcounter{tocdepth}{2}}

\paragraph{Independent variables.}
In both studies, we manipulate three independent variables. The first, \textit{AI support condition}, is a between-subjects variable, meaning that each participant is assigned to only one of its four conditions. The other two variables, \textit{item's ground-truth label} and \textit{model's classification accuracy}, are within-subjects variables, meaning that each participant encounters all levels of these variables across the items they classify.

\begin{figure*}[t]
\centering
\includegraphics[height=8.5cm]{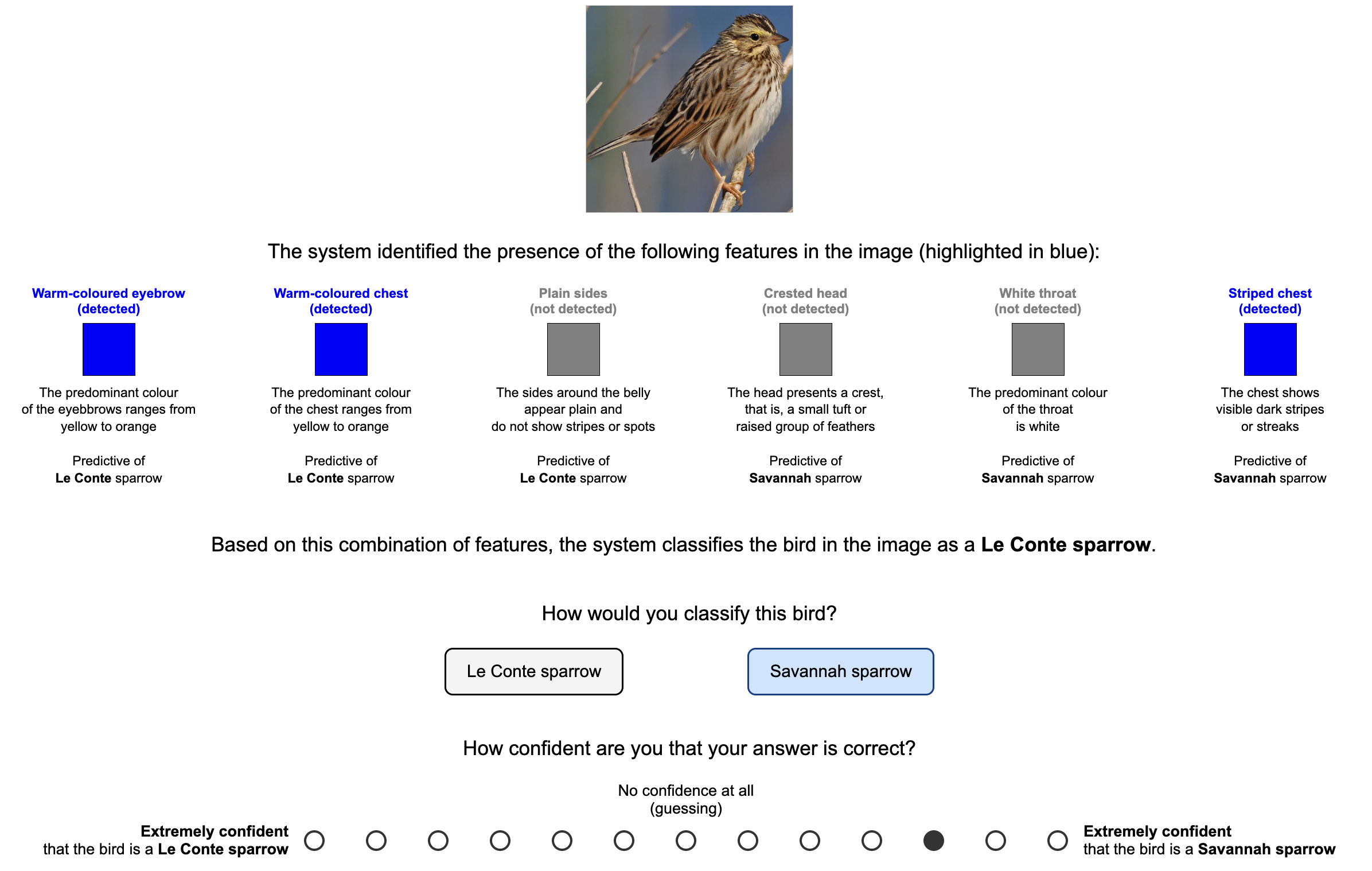}
\caption{Interface presented in the NIC and IC conditions of Study 2. The
interface presented in the LO and NS conditions was largely identical, except that the concept information and, in the
latter case, also the model prediction, were removed (see \SM for the interface presented in Study 1).}
\label{interface}
\end{figure*}

\begin{enumerate}
    \item \textbf{AI support condition}. We randomly assign participants to one of four conditions:
    \begin{enumerate}
        \item \textit{No support - NS}: Participants perform the classification task without receiving any support from the CBM.
        \item \textit{Label only - LO}: Participants perform the task receiving only the labels predicted by the CBM, with no access to the predicted concepts, acting as a non-interpretable decision support system.
        \item \textit{Non-interactive concepts - NIC}: Participants receive both the predicted label and the set of concepts detected and not detected by the CBM, but cannot modify the values of the concepts.
        \item \textit{Interactive concepts - IC}: Participants receive the same information as in the NIC condition, but can also modify concept values.
    \end{enumerate}

    In the NIC and IC conditions, all six concepts are displayed as boxes colored in blue when the concept is detected and in gray when it is not (see \cref{interface}).
    
    \item \textbf{Item's ground truth label}. In both studies, we ask participants to classify 10 items, evenly divided between the two task classes (fraudulent and legitimate in Study 1, Le Conte's sparrow and Savannah sparrow in Study 2).
    \item \textbf{Model's classification accuracy}. We stratify the random sampling of items so that participants are presented with eight items that the model classifies correctly (four from each class) and two that it classifies incorrectly (one from each class). Consequently, participants in the three AI-supported conditions receive predictions from a model with an accuracy of 80\%. This approximates the original test accuracy of the CBMs while ensuring that participants encounter some incorrectly classified items, which are extremely relevant to consider for evaluating CBMs as decision-support systems.
\end{enumerate}

\paragraph{Metrics.}

We consider the following metrics:
\begin{enumerate}
    \item \textbf{Participants' accuracy}: the proportion of participants’ classifications that match the ground-truth labels.
    \item \textbf{Participants' classification confidence}: confidence is rated on a 13-point scale ranging from 1 (“Extremely confident that the item belongs to Class 1”) to 13 (“Extremely confident that the item belongs to Class 2”), with 7 indicating no confidence for either class (i.e., a guess).
    \item \textbf{Participants' intervention} (IC condition only): the number of times participants intervene on concept values.
    \item \textbf{Participants' trust in the model} (AI-supported conditions only): the average rating across eight trust scale items adapted from \citet{DBLP:journals/fcomp/HoffmanMKL23}, see \SM.
\end{enumerate}

\paragraph{Procedure.}

We recruit participants through Prolific and randomly assign them to one of the four conditions. They first receive instructions on how to perform the task and then complete two practice trials (tailored to their assigned condition). To encourage attentive performance, we inform participants within each condition that the five most accurate participants would receive a £5.00 bonus in addition to their Prolific payment.

Participants then complete the main task, which consists of classifying 10 items. In both studies, these items are randomly drawn from a subset of the test set on which the CBM had been deployed, and they are presented to participants in a random order. 
For each item, participants select the class to which they believe the item belongs and rate their confidence in their classification. Importantly, no feedback is provided on the accuracy of their responses. This both reflects realistic decision-support settings, in which the correctness of a classification may not be immediately verifiable, and limits learning across trials.

To reduce inattentive responding, participants can proceed to the next item only if their selected class is consistent with their confidence rating (i.e., they cannot select one class while expressing confidence that the item belongs to the other). We include two attention checks and record the number of times participants switch away from the experiment browser tab. After classifying all items, participants in the AI-supported conditions complete the trust questionnaire. Finally, all participants report their familiarity with AI systems.

\paragraph{Analyses.}
We analyze the results as follows.

\begin{itemize}
    \item \textbf{Classification accuracy:} We fit two logistic mixed-effects regression models\footnote{Mixed-effects regression models extend standard regression by including random effects that account for the non-independence of repeated observations (e.g., multiple responses from the same participant or to the same item), yielding valid inference despite correlated observations (for an overview, see \citealp{brown2021introduction}).}.
    The first includes only AI support condition as a fixed effect to assess the overall impact of the four experimental conditions on participants' accuracy and is followed by six pairwise post-hoc comparisons. The second also includes the item's ground-truth label, the system's classification accuracy, and all interactions among these variables, and is followed by 12 pairwise post-hoc comparisons.

    \item \textbf{Confidence ratings:} We recode confidence ratings on a scale from 0 to 6, with higher values indicating greater confidence that the classification was correct, regardless of the class assigned. We then analyze these ratings using an ordinal mixed-effects regression model, with AI support condition as a fixed effect, followed by six post-hoc comparisons.
        
    \item \textbf{Intervention behavior:} We fit two logistic mixed-effects regression models focused only on participants in the IC condition. The first model assesses, at the concept level, whether participants are more likely to modify a concept when its detected value differs from its ground-truth annotation. The second assesses whether intervening at least once while evaluating an item improves its final classification accuracy.
     
    \item \textbf{Trust:} We compare average trust across the three AI-supported conditions using a Kruskal–Wallis test. We also conduct separate Kruskal-Wallis tests for each of the eight questionnaire items, treating these as a family of eight tests. Significant results are followed by pairwise Wilcoxon rank-sum tests (three comparisons).

\end{itemize}

All mixed-effects models described above include random intercepts for participants and items. The model on participants' propensity to intervene on concepts also includes random intercepts for concepts. When multiple hypothesis tests are performed (e.g., pairwise post-hoc comparisons), we apply Bonferroni corrections to the \textit{p} values to control the family-wise error rate at $\alpha = .05$. We report corrected \textit{p} values throughout. (See \SM for the full analyses report.)

\paragraph{Sample characteristics.}  

We refer the reader to the \SM for details on sample size determination and data collection. Here, we note that we exclude participants who fail one or more attention checks or switch away from the experiment browser tab more than three times during the study.

For Study 1, we collected a total of 401 participants and, following exclusions, the final sample consisted of 363 participants (female = 49\%, \textit{M}\textsubscript{\textit{age}} = 40.15, \textit{SD}\textsubscript{\textit{age}} = 12.92). For Study 2, we collected a total of 551 participants and, following exclusions, the final sample consisted of 342 participants (female = 55\%, \textit{M}\textsubscript{\textit{age}} = 38.05, \textit{SD}\textsubscript{\textit{age}} = 11.92; see \SM for a discussion of the differing exclusion rates in Studies 1 and 2). A chi-squared test of independence indicates that experience with AI systems does not differ significantly across conditions in both studies (\textit{p} = .402 and \textit{p} = .312, respectively).

\paragraph{Research questions.}
We explore the following three research questions:

\begin{itemize}
    \item \textbf{Q1}: Do CBMs, and in particular their interactive component, improve the accuracy of human-AI teams in classification tasks? If so, do these benefits depend on the correctness of the CBM's prediction?
    \item \textbf{Q2}: How do users actually interact with CBMs? In particular, do they intervene primarily when concept detection is incorrect, consistent with the intended use of CBMs?
    \item \textbf{Q3}: Do CBMs improve users' confidence in their classifications and their trust in the model compared with a non-interpretable AI system?
\end{itemize}

\addtocontents{toc}{\protect\setcounter{tocdepth}{0}}
\section{Study 1 – Fuzzy concepts in a familiar task}
\addtocontents{toc}{\protect\setcounter{tocdepth}{2}}

For Study 1, we select a classification task with which participants are presumably somewhat familiar and which we therefore expect them to approach with a certain degree of confidence. However, the concepts used by the CBM are not entirely clear-cut and involve some subjectivity in determining whether they are present in the item being classified. Specifically, participants are presented, one at a time, with the subject line and body of 10 emails and asked to classify each email as either legitimate or fraudulent (i.e., a phishing attempt) based on emotional and goal-oriented cues.

Among the test set of the \Emails dataset, we select six emails for which the model produces an incorrect prediction (three phishing and three legitimate emails) and 24 emails for which the model produces the correct prediction (12 phishing and 12 legitimate emails). Items are selected to maximize topic diversity and classification difficulty (see \SM for details on item selection). The final item pool comprises 30 emails, from which we randomly extract the 10 items to be presented to each participant.\footnote{Following data collection, observations corresponding to one of the selected emails (91 out of 3630 total observations) were excluded because of incorrect concept activations due to a human error. However, retaining these observations did not alter the results (see \SM for details).}

\begin{figure*}[t]
\centering
\includegraphics[height=4.8cm]{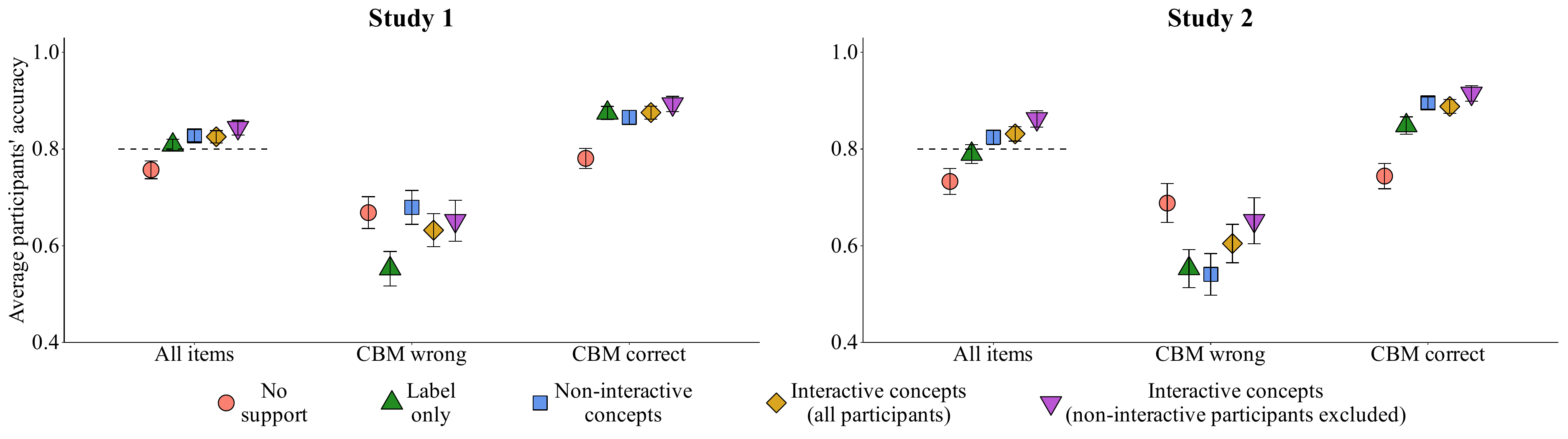}
\caption{Participants' classification accuracy by condition, overall and by CBM classification accuracy, for the full sample and after excluding IC participants who never interacted with the concepts (error bars represent standard errors). The dashed line indicates CBM's accuracy for the selected items.}
\label{accuracy}
\end{figure*}

\addtocontents{toc}{\protect\setcounter{tocdepth}{0}}
\subsection{Results}
\addtocontents{toc}{\protect\setcounter{tocdepth}{2}}

\paragraph{R1: CBMs provide limited accuracy benefits over label-only support.} 

As shown in \cref{accuracy}, participants' overall classification accuracy is significantly lower in the NS condition (76\%) than in the LO (81\%, \textit{p} = .043), NIC (83\%, \textit{p} < .001), and IC (83\%, \textit{p} = .002) conditions, whereas the three AI-supported conditions do not differ significantly from one another (\textit{p}s = 1). A more detailed analysis reveals a significant interaction between AI-supported condition and the CBM's classification accuracy (\textit{p} = .002): for items that the CBM classifies incorrectly, participants in the NIC condition (68\%) are more accurate than those in the LO condition (55\%, \textit{p} = .047).

\paragraph{R2: Participants intervene mainly when concept detection is incorrect, but without accuracy gains.}

Participants in the IC condition intervened an average of 6.87 times (\textit{SD} = 7.69) across the 10 items. Interventions are more frequent when the detected concept differs from the ground-truth annotation (22\%) than when the two match (9\%, \textit{p} < .001). Yet, intervening on concept values at least once when evaluating an item does not significantly improve classification accuracy (accuracy in trials with and without interventions is equal to 82\% and 83\%, respectively; \textit{p} = .248).

It is also worth noting that 31 of the 87 participants in the IC condition (36\%) never intervened on the concepts. Excluding these participants, however, does not alter the results for classification accuracy.

\paragraph{R3: CBMs do not affect participants' confidence in their classifications,  but trust in the model may be slightly reduced for interactive CBMs.}

Overall, no significant differences are present across the four experimental conditions in terms of participants' classification confidence (\textit{p} = .588). Excluding non-interactive participants in the IC condition does not alter these results.

Also, no significant differences across the three AI-supported conditions are observed in participants' self-reported trust in the model, neither for the overall trust index (\textit{p} = .187) nor for any of the individual items (lowest \textit{p} value equal to .519). However, when non-interactive participants in the IC condition are excluded, ratings for one questionnaire item ("The system can perform the task better than a novice human user") result to be significantly different across conditions (\textit{p} = .044). Specifically, participants in the IC condition present lower ratings than those in the LO condition (medians equal to 3.5 and 4.0, respectively; \textit{p} = .007), suggesting that actively interacting with concepts may have partly undermined trust in the model's capabilities.

\addtocontents{toc}{\protect\setcounter{tocdepth}{0}}
\section{Study 2 – Well-defined concepts in an unfamiliar task}
\addtocontents{toc}{\protect\setcounter{tocdepth}{2}}

In Study 2, we investigate whether the effectiveness of CBMs would emerge more clearly in a less familiar task, in which participants are therefore expected to feel relatively less confident in their classifications, yet the concepts used by the model refer to features that are more readily observable in the items. Specifically, we ask participants to classify images of birds as belonging to one of two sparrow species based on a set of visual cues (e.g., breast color).

We select a pool of 20 images from the original \Cub test set, of which four are incorrectly and 16 correctly classified by the CBM. Each subgroup contains an equal number of Le Conte's and Savannah sparrows (see the \SM for details of the image selection criteria).\footnote{In the \Cub dataset, concept annotations are assigned at the species level rather than being manually verified for each individual image. As a result, some images are annotated as containing concepts that are not actually visible (e.g., a bird photographed from behind may still be labeled as having a striped breast). This may confuse users when attempting to understand how the model inferred the presence or absence of those concepts from the input image. To avoid this issue, we select images in which all concepts used in the CBM bottleneck are clearly visible and manually verify the ground-truth annotations for these concepts (see \SM for details).} For each participant, we randomly sample 10 items from the pool to present during the experiment.

\addtocontents{toc}{\protect\setcounter{tocdepth}{0}}
\subsection{Results}
\addtocontents{toc}{\protect\setcounter{tocdepth}{2}}

\paragraph{R1: CBMs improve users’ accuracy compared with no support, whereas label-only support does not.} 

As it can be seen in \cref{accuracy}, participants' overall classification accuracy is significantly higher in the NIC (82\%, \textit{p} = .016) and IC (83\%, \textit{p} = .005) conditions than in the NS condition (73\%), whereas accuracy in the LO condition (79\%) does not differ significantly from that in NS (\textit{p} = .512). Accuracy in the NIC and IC conditions also does not differ significantly from that in LO (\textit{p} = 1 and \textit{p} = .637, respectively). A more detailed analysis again reveals a significant interaction between condition and the CBM's classification accuracy (\textit{p} < .001). The only significant contrasts indicate that, when the CBM's predictions are correct, participants are more accurate in the AI-supported conditions (LO: 85\%, \textit{p} = .009; NIC: 90\%, \textit{p} < .001; IC: 89\%, \textit{p} < .001) than in the NS condition (74\%).

\paragraph{R2: Users who interact with the CBM outperform users with label-only support.}

Participants in the IC condition intervened an average of 9.62 times (SD = 11.02) across the 10 items. Replicating and strengthening the pattern observed in Study 1, participants interact with concepts more often when the detected concept differs from the ground-truth annotation (33\%) than  when the two match (5\%, \textit{p} < .001). 
Moreover, classification accuracy is significantly higher on trials in which participants intervene at least once than on those with no interventions (88\% vs. 79\%, \textit{p} < .001). 

As in Study 1, 30 of the 86 participants (35\%) in the IC condition never interacted with the concepts. In this case, however, excluding these participants increases accuracy in this condition to 86\%, making it significantly higher than that in the LO condition (\textit{p} = .049). This improvement is driven primarily by higher classification accuracy on items for which the CBM predicts an incorrect label, indicating participants' greater resistance to model's errors (65\% vs. 60\% accuracy, respectively).
A smaller improvement in accuracy is present also for items for which the CBM predicts the correct label (91\% vs. 89\%), plausibly because participants are more likely to follow the model’s correct suggestions than to override them with their own incorrect predictions.

\paragraph{R3: Interactive CBM increases participants' confidence in their classifications, but may slightly reduce their trust in the model.}

Overall, participants' confidence in their classifications differs across conditions (\textit{p} = .025): it is higher in the IC than in the NS condition (medians equal to 4 and 3, respectively; \textit{p} = .025), whereas all other pairwise comparisons are not significant (lowest \textit{p} value equal to .134).

Finally, no significant differences across the three AI-supported conditions are observed in participants' trust in the model, neither for the overall trust index (\textit{p} = .076) nor for any individual questionnaire item (lowest \textit{p} value equal to .058). However, as in Study 1, when non-interactive participants in the IC condition are excluded, ratings for the item "The system can perform the task better than a novice human user" differ across conditions (\textit{p} = .048), with participants in the IC condition giving lower ratings (median = 4) than those in the LO condition (median = 5; \textit{p} = .005).

\addtocontents{toc}{\protect\setcounter{tocdepth}{0}}
\section{Related Work}
\addtocontents{toc}{\protect\setcounter{tocdepth}{2}}

\paragraph{Concept Bottleneck Models.}
Recent work has addressed numerous aspects of CBMs \citep{DBLP:journals/tmlr/KnabSBKSSSS26}, 
including obtaining concepts with a human-aligned semantics \citep{DBLP:conf/nips/MarconatoPT22, DBLP:conf/nips/HavasiPD22, 
DBLP:journals/ml/DeboleBGPTM26, colamonacoprototype, zarlengaposition}, 
designing bottlenecks to model sufficient task statistics \citep{DBLP:journals/access/SawadaN22, kalampalikis2025towards}, integrating interpretable (beyond linear) task predictors \citep{debot2024interpretable, stammer2024neural,  de2026causally},
defining effective interventional policies and how to involve humans~\citep{DBLP:conf/icml/SteinmannSFK24, zarlenga2025avoiding, DBLP:journals/corr/abs-2503-16199}, and how to extract concepts leveraging vision-language models~\citep{DBLP:conf/cvpr/YangPZJCY23, DBLP:conf/iclr/OikarinenDNW23, DBLP:conf/nips/SrivastavaYW24, debole2026concepts}.
It is worth noting that these works mainly target the structural components and key properties of CBMs, whereas our work tests an orthogonal, yet central aspect of CBMs: their efficacy in hybrid decision-making.

\paragraph{Concept-based XAI user studies.}

Existing user studies suggest that concept-based explanations are judged as adequate and informative \citep{DBLP:conf/nips/BhallaOSCL24, DBLP:conf/emnlp/RajagopalBHT21}, easy to interpret \citep{DBLP:journals/tmlr/ByrmanKKU25, DBLP:journals/tvcg/HuangMKB23, DBLP:journals/tmlr/SchrodiSAB25}, and helpful for predicting and, to some extent, critically evaluating model outputs \citep{DBLP:conf/iclr/AdebayoMAK22, DBLP:conf/nips/DubeyR022, DBLP:conf/cvpr/RamaswamyKFR23}.
The works by \citet{DBLP:conf/iclr/BontempelliTTGP23,DBLP:journals/corr/abs-2506-05533} have focused on interactions with user to revise incorrect concept prototypical predictions. 
However, most of these studies do not evaluate metrics such as human–AI team performance or users’ confidence in their judgments, which are crucial when assessing the practical benefits of an explainability approach \citep{DBLP:conf/iui/BucincaLGG20, DBLP:journals/corr/abs-2406-08271}. Two notable exceptions in this regard are the works by \citet{DBLP:conf/nips/DasCK23} and \citet{DBLP:conf/xai/FurbyCBP25}. However, also in these studies important aspects are not assessed, and methodological shortcomings limit the robustness of the results.

In particular, \citet{DBLP:conf/nips/DasCK23} found that concept-based explanations (but not specifically CBMs) improved user performance across two game-like tasks. However, several aspects limit the relevance of this study to our work. First, the tasks involved sequential decision-making rather than the classification problems for which CBMs are typically designed.
Second, participants played the game unaided twice, with a supported session in between. Performance was assessed in terms of improvement from the first to the second unaided phase, making the study more informative about concept-based explanations as training tools than as decision-support systems. Third, participants could not intervene on the concepts, leaving the interactive component of concept-based explanations unexplored. Finally, the small sample sizes (15 participants per condition) limited the statistical robustness of the effects.

\citet{DBLP:conf/xai/FurbyCBP25} specifically investigated CBMs, allowing some groups of participants to intervene on concept values. Their lay-user study provided some evidence that interaction with concepts can improve human-AI team performance, particularly when model accuracy is relatively low. However, the sample size remained limited (13 participants per condition), the Blackjack task employed was not representative of typical CBM applications, and the interface required participants to modify inherently binary concepts (e.g., whether the dealer's cards value was equal to 10) using continuous sliders. This may have made intermediate values difficult to interpret and introduced noise into the results. Moreover, the study did not include a no-support condition, leaving open the question of how much AI support itself improved participants' performance.  Their expert study addressed some of these concerns by using a realistic image-classification task with participants who had expertise in dermatology. However, it also lacked a no-support condition and, in addition, did not include either a label-only condition or a non-interactive concept condition, making it impossible to disentangle the effects of AI support, concept explanations, and concept intervention.

\addtocontents{toc}{\protect\setcounter{tocdepth}{0}}
\section{Conclusions}
\addtocontents{toc}{\protect\setcounter{tocdepth}{2}}

The results of our studies indicate that CBMs, and particularly their interactive component, can improve human–AI team performance beyond that obtained with non-interpretable AI support or by humans acting alone.

However, our findings suggest that these benefits may be more likely to emerge under the following conditions:
\begin{enumerate}
\item The task should be one in which users are relatively uncertain about the correct classification. Indeed, users who are fairly - or even highly - confident in their own classifications understandably have less reason to rely on the explanations provided by the CBM.
\item By contrast, the concepts used by the CBM should be as objectively grounded as possible and easy for users to understand. Indeed, fuzzy concepts that are open to subjective interpretation may obscure the rationale underlying the model’s classifications, thereby reducing users’ trust in the model.
\item Users must be willing to engage with the interactive component of the CBM. In the IC condition of our studies, approximately one third of participants never modified any concept values. This suggests that merely providing users with the opportunity to interact with concept-based explanations is not sufficient for them to make use of this feature, which appears crucial for CBMs to deliver their full benefits.\footnote{The stronger performance observed in Study 2 after excluding non-interacting participants may partly reflect self-selection, as participants who chose to interact may have been more motivated or engaged. However, self-selection is unlikely to fully account for the findings, because applying the same exclusion criterion in Study 1 did not yield a comparable improvement in performance.}
\end{enumerate}

Two limitations of our work should be acknowledged. First, we focused exclusively on binary classification tasks, whereas the effectiveness of CBMs in multi-class settings remains to be explored. On the one hand, users may be less confident in their classifications, thereby increasing the potential value of decision-support systems based on CBMs. On the other hand, multi-class tasks typically require a larger number of concepts, which may substantially increase users' cognitive load. Approaches that flexibly select subsets of concepts for presentation to users 
may alleviate users' cognitive burden without compromising the benefits of concept-based reasoning~\cite{barker2023selective, DBLP:journals/tmlr/SchrodiSAB25}.
Second, we did not systematically investigate the impact of concept-detection accuracy. A CBM that achieves good classification performance while frequently misidentifying concepts may undermine users' trust in the model, leading them to discount its predictions and thereby reducing their potential benefits (see the \SM for exploratory analyses that are consistent with this possibility). More generally, a major problem with CBMs is that concept annotations are not always reliable, which may substantially compromise their considerable potential as decision-support systems (see also footnote 2).

\bibliography{bibliography}

\appendix

\newpage

\begin{center}
    \Large\textbf{Supplementary Materials}
\end{center}

\tableofcontents

\newpage

\section*{CBM specifics}
\addcontentsline{toc}{section}{CBM specifics}
\label{appendix:cmb_specifics}
\bigskip

The CBM follows the independent training variant~\cite{DBLP:conf/icml/KohNTMPKL20}, in which the concept extractor and task predictor are trained separately.

\paragraph{Concept Extractor.}
\label{appendix:cbm_concept_extractor}
The concept extractor learns a mapping from the input $\mathbf{x}$ to a vector of $n_c$  concept activations.
Implementation-wise, $g$ consists of a frozen neural encoder, which produces text or image embeddings, followed by $n_c$ independent binary SVM classifiers, one for each concept. 
Each SVM is trained using the \texttt{sklearn} implementation\footnote{\url{https://scikit-learn.org/stable/modules/generated/sklearn.svm.SVC.html}} with hyperparameters \verb|kernel=rbf|, \verb|C=1.0|, and \verb|class_weight=balanced|.
The output of each SVM is a decision score (logit) representing the presence or absence of the corresponding concept.

For the \Cub dataset, we use the pretrained CLIP ViT-L14 vision encoder \citep{DBLP:conf/icml/RadfordKHRGASAM21} as the backbone of the concept extractor. 
For the \Emails dataset, we instead use a pretrained sentence transformer \citep{DBLP:conf/emnlp/ReimersG19}, based on \citet{DBLP:conf/nips/WangW0B0020}, as the encoder.
This encoder choice is the only architectural difference between the two settings, as the image and text modalities require different embedding models. 
Once the embeddings are extracted, the subsequent concept extraction pipeline is identical across both datasets.

\paragraph{Task Predictor.}
The task predictor $f$ maps concept vectors to the task prediction.
Ground-truth concept annotations are rescaled to $\{-1,+1\}^{n_c}$ (negative and positive classes, respectively), following the convention used in support vector machines and other margin-based classifiers. This representation provides a symmetric encoding of the two classes and is consistent with the signed outputs produced by the SVM concept predictors~\citep{DBLP:journals/ml/CortesV95}.
In line with works on CBMs, we  implement $f$ as a logistic regression model, using weights $\mathbf w \in \mathbb R^{n_c}$ and with no bias term.
Notice that this is equivalent to using concept activations in the $[0,1]^{n_c}$ interval but it allows the model to explicitly take into account the absence of concepts in the input, e.g., the absence of fear-related concepts in an email may provide evidence that the email is legitimate.

\paragraph{Test-time predictions.}
\label{appendix:cbm_inference}
At test time, the input $\mathbf{x}$ is first passed through the concept extractor $g$ to obtain a vector of concept logits, we then apply a $\tanh(\cdot)$ activation function to map them to shifted concept activations $[-1,1]^{n_c}$. 
Afterwards, the concept activations are passed to the task predictor $f$, which outputs the probability of the positive class. The positive class (legitimate for \Emails and Savannah sparrow for \Cub) is predicted if the probability is equal or above $0.5$.

\paragraph{Human Intervention.}
\label{appendix:cbm_intervention}
We allow human intervention on concepts during inference by enabling users to modify the activation of any concept to reflect their belief about whether that concept is present. 
In our implementation,
users can only toggle a concept between the active and inactive states, without assigning intermediate confidence values (e.g., a concept cannot be set to a high probability; it can only be activated or deactivated).
Although continuous interventions would more closely reflect the internal representation of concepts in the model, they could introduce unnecessary variability by requiring participants to interpret the meaning of intermediate activation values. This concern is particularly relevant for Study 2, where concepts correspond to visually identifiable attributes (e.g., the presence of a striped chest or a white throat), for which assigning a partial degree of presence could be counterintuitive. 
Therefore, restricting interventions to binary decisions makes the interface easier to understand, reduce potential ambiguity in participants' responses, and ensures a consistent intervention mechanism across both user studies.

Specifically, when a user intervenes on a concept, its activation is replaced with the opposite polarity: 
interventions on positive activations transform the concept value $-1$, while interventions on negative activations to $1$. 
Clicking the concept again restores its original predicted value. In this way, the interface allows the human to override the model's concept predictions and adjust its reasoning whenever a predicted concept differs from their belief.
E.g., if the model predicts $c_\text{round}= -0.5$, the only allowed intervention on it is $(c_\text{round} \leftarrow +1)$, and a second click by the user would restore $c_\text{round} = -0.5$.

\newpage
\section*{Datasets}
\addcontentsline{toc}{section}{Datasets}
\label{appendix:datasets}
\bigskip

\paragraph{\Emails}
The training split contains 1,064 emails after cleaning the dataset (see section Study 1 - Concepts and email selection below). To increase the amount of training data available for the \textbf{concept predictor}, we retain the emails belonging to the \textit{Spam} class. However, when training the \textbf{task predictor}, we remove the 513 emails belonging to the \textit{Spam} class.
\paragraph{\Cub}
The training dataset contains 4,796 images available across 200 classes. To train the \textbf{concept predictor}, we use all images except those belonging to the \textit{Le Conte's Sparrow} and \textit{Savannah Sparrow} classes, leaving us with 4,745 images. Since each image is associated with 6 concept annotations, the total number of concept annotations is 28,470.
To train the \textbf{task predictor}, we instead use only the samples belonging to the \textit{Le Conte's Sparrow} and \textit{Savannah Sparrow} classes, as our task consists of binary classification.

We summarize in \cref{tab:train-stats} useful statistics regarding the \Emails and \Cub datasets.

\begin{table}[hb]
\centering

\begin{subtable}[t]{0.48\textwidth}
\centering
\begin{tabular*}{\linewidth}{@{\extracolsep{\fill}}lrr@{}}
\toprule
Concept & Percentage & Occurrence \\
\midrule
Fear+Authority  & 40.60 & 432 \\
Urgency         & 55.08 & 586 \\
Curiosity       & 46.05 & 490 \\
Neutral         & 29.14 & 310 \\
Reply           &  5.73 &  61 \\
Open attachment &  4.70 &  50 \\
\bottomrule
\end{tabular*}
\subcaption{\Emails dataset statistics. It is used to train the concept predictor, and consists in the 1064 training images, each annotated with 6 concepts.}
\end{subtable}
\hfill
\begin{subtable}[t]{0.48\textwidth}
\centering
\begin{tabular*}{\linewidth}{@{\extracolsep{\fill}}lrr@{}}
\toprule
Class & Percentage & Occurrence \\
\midrule
Phishing (0) & 66.42 & 366 \\
Valid (1)    & 33.58 & 185 \\
\midrule
\textbf{Total} & \textbf{100} & \textbf{551} \\
\bottomrule
\end{tabular*}
\subcaption{\Emails dataset statistics, used to train the task predictor. It is obtained by removing the emails belonging to the "Spam" class from the dataset.}
\end{subtable}

\vspace{1em}

\begin{subtable}[t]{0.48\textwidth}
\centering
\begin{tabular*}{\linewidth}{@{\extracolsep{\fill}}lrr@{}}
\toprule
Concept & Percentage & Occurrence\\
\midrule
striped breast &  6.81 &   323 \\
buff breast    & 11.53 &   547 \\
white throat   & 36.97 & 1,754 \\
buff nape      &  8.01 &   380 \\
solid belly    & 75.13 & 3,565 \\
brown crown    & 13.78 &   654 \\
\bottomrule
\end{tabular*}
\subcaption{\Cub dataset statistics. It is used to train the concept predictor, and consists in the 4745 training images (after removing those belonging to the \textit{Le Conte's} and \textit{Savannah sparrow} classes), each annotated with 6 concepts.}
\end{subtable}
\hfill
\begin{subtable}[t]{0.48\textwidth}
\centering
\begin{tabular*}{\linewidth}{@{\extracolsep{\fill}}lrr@{}}
\toprule
Class & Percentage & Occurrence \\
\midrule
Le Conte's Sparrow (123) & 49.02 & 25 \\
Savannah Sparrow (126)   & 50.98 & 26 \\
\midrule
\textbf{Total} & \textbf{100} & \textbf{51} \\
\bottomrule
\end{tabular*}
\subcaption{\Cub dataset statistics, used to train the task predictor. It is obtained by selecting only the training images belonging to the \textit{Le Conte's} and \textit{Savannah sparrow}.}
\end{subtable}

\caption{Statistics of the training datasets.}
\label{tab:train-stats}
\end{table}

\newpage
\section*{A priori power analysis and data collection}
\addcontentsline{toc}{section}{A priori power analysis and data collection}
\bigskip

We conducted an a priori power analysis using a simulation-based approach \citep{green2016simr, kumle2021estimating} to estimate the sample size required to detect a small effect of the three-way interaction between AI-support condition, the item’s ground-truth label, and the model’s classification accuracy on participants’ classification accuracy. The analysis indicated a required sample size of 340 participants (i.e., 85 for each AI support condition) to achieve 82\% of statistical power. Participants were recruited in batches. After each batch, we assessed only whether participants met the predetermined exclusion criteria (see main paper); no analyses of the outcome variables or effect sizes were performed during data collection. Recruitment continued until the final sample comprised at least 85 eligible participants in each experimental condition. Inclusion criteria required participants to be native English speakers from the UK and to have a Prolific approval rate above 98\%. Participants received compensation equal to £1.30.

As briefly mentioned in the main paper, Study 2 presented a much higher exclusion rate (209 participants out of a total sample of 551, 38\%) than Study 1 (38 participants out of 401, 9\%). This was caused in particular by a large number of participants (196) failing at least one attention check in Study 2. This may partly reflect lower engagement, as participants were asked to classify unfamiliar bird species. Exclusions were also disproportionately frequent among participants in the \textit{No-support} condition (130 of the 196 participants who failed at least one attention check), possibly because, unlike those in the AI-supported conditions, they were not accustomed to receiving textual information on the interface and may therefore have overlooked the instructions embedded in the attention checks. Exclusions were indeed substantially lower in the AI-supported conditions (\textit{Label-only} = 38, \textit{Non-interactive concepts} = 19, \textit{Interactive concepts} = 9).

Due to this high exclusion rate, particularly in the \textit{No-support} condition, proceeding with fully random assignment throughout data collection would have resulted not only in an extremely expensive recruitment process but also in highly imbalanced numbers of participants across the experimental conditions, which could have reduced the precision and robustness of the statistical analyses (e.g., \citealt{harrison2018brief}). Therefore, after the first 400 participants, who were assigned completely at random to the experimental conditions, the remaining 151 participants were assigned to specific conditions until each condition retained at least 85 eligible participants after exclusions. However, restricting the analyses to the initial 400 fully randomized participants did not alter the results.

\newpage

\section*{Trust scale}
\addcontentsline{toc}{section}{Trust scale}
\bigskip

The trust scale used in Studies 1 and 2 was adapted from \citet{DBLP:journals/fcomp/HoffmanMKL23}. In particular, three items were reworded compared to the original scale to invert their polarity, a common practice to reduce the risk of acquiescence bias (i.e., participants' tendency to express agreement with the items of a questionnaire). The items, presented in randomized order, were the following:
\begin{itemize}
    \item I am confident in the system. I feel that it works well.
    \item The outputs of the system are very unpredictable [reworded; original: The outputs of the system are very predictable].
    \item The system is very reliable. I can count on it to be correct all the time.
    \item I feel safe that when I rely on the system I will get the right answers.
    \item The system is inefficient in that it works very slowly [reworded; original: The system is efficient in that it works very quickly].
    \item I am wary of the system.
    \item The system can perform the task better than a novice human user.
    \item I do not like using the system for decision making [reworded; original: I like using the system for decision making].
\end{itemize}

Each item was measured using a 7-point Likert scale (extremes: I strongly disagree - I strongly agree)

\bigskip

\section*{Previous experience with AI systems}
\addcontentsline{toc}{section}{Previous experience with AI systems}
\bigskip

At the end of both Studies 1 and 2, participants were asked to select the option among the following four that best described their level of experience with AI systems (see Table S1 for the distribution of answers in the final samples of the two user studies):

\begin{itemize}
    \item \textbf{Option 1}: I have little or no experience with AI systems and limited or no understanding of how they work.
    \item \textbf{Option 2}: I use AI systems occasionally but have not a clear understanding of how they function.
    \item \textbf{Option 3}: I use AI systems and have studied how they work (e.g., through courses, online classes, or self‑study).
    \item \textbf{Option 4}: I develop or build AI systems as part of my work or personal projects.
\end{itemize}

\begin{table}[!h]
\centering
\captionsetup{skip=5pt}
\scriptsize
\begin{tabular}{lcccc}
\toprule
\textbf{Condition} & \textbf{Option 1} & \textbf{Option 2} & \textbf{Option 3} & \textbf{Option 4} \\
\midrule

\multicolumn{5}{l}{\textbf{Study 1}}\\
\midrule

No Support &
7 (7.9\%) &
47 (52.8\%) &
31 (34.8\%) &
4 (4.5\%) \\

Label Only &
2 (2.1\%) &
56 (58.9\%) &
35 (36.8\%) &
2 (2.1\%) \\

Non-Interactive &
4 (4.3\%) &
51 (55.4\%) &
34 (37.0\%) &
3 (3.3\%) \\

Interactive &
2 (2.3\%) &
54 (62.1\%) &
31 (35.6\%) &
0 (0.0\%) \\

\midrule
\multicolumn{5}{l}{\textbf{Study 2}}\\
\midrule

No Support &
5 (5.9\%) &
49 (57.6\%) &
30 (35.3\%) &
1 (1.2\%) \\

Label Only &
5 (5.9\%) &
49 (57.6\%) &
31 (36.5\%) &
0 (0.0\%) \\

Non-Interactive &
5 (5.9\%) &
51 (59.3\%) &
28 (32.6\%) &
2 (2.3\%) \\

Interactive &
4 (4.7\%) &
52 (60.5\%) &
24 (27.9\%) &
6 (7.0\%) \\

\bottomrule
\end{tabular}
\caption{Distribution of participants' prior AI experience across experimental conditions in Studies 1 and 2. Counts are reported, with row percentages in parentheses.}
\label{tab:experience}

\end{table}

\newpage

\section*{Highest and lowest confidence ratings in Studies 1 and 2}
\addcontentsline{toc}{section}{Checking maximium and minimum confidence ratings in Studies 1 and 2}
\bigskip
We tested whether the frequencies of the highest (i.e., 6) and lowest (i.e., 0) recoded confidence ratings followed the expected pattern given the characteristics of the two tasks (relatively high confidence even in the NS condition in Study 1, and lower confidence in the NS condition than in the AI-supported conditions in Study 2). To this end, we fitted two Poisson regression models, followed by six post-hoc comparisons, to assess whether the counts for the two ratings of interest differed across experimental conditions. Reported \textit{p} values were Bonferroni corrected to maintain the family-wise error rate at $\alpha = .05$

\bigskip
\textbf{Study 1 - Email classification} As we expected for this task, the highest confidence rating was relatively common, accounting for 25\% of all responses, and its frequency was not different across the four experimental conditions 
(NS: \textit{M} = 2.79, \textit{SD} = 2.66;  
LO: \textit{M} = 2.49, \textit{SD} = 2.42; 
NIC: \textit{M} = 2.23, \textit{SD} = 2.56; 
IC: \textit{M} = 2.40, \textit{SD} = 2.32; 
\textit{p} = .117). In contrast, the lowest confidence rating was rare, accounting for 3\% of all responses, and its frequency again did not differ across conditions 
(NS: \textit{M} = 0.29, \textit{SD} = 0.76;  
LO: \textit{M} = 0.35, \textit{SD} = 0.73; 
NIC: \textit{M} = 0.35, \textit{SD} = 1.23; 
IC: \textit{M} = 0.24, \textit{SD} = 0.78; 
\textit{p} = .504).

\paragraph{Study 1 -- Highest confidence rating -- AI-support condition}
\textcolor{white}{[parkour]}

\begin{table}[!h]
\centering
\captionsetup{skip=5pt}
\begin{tabular}{lcccc}
\toprule

\multicolumn{5}{l}{\textbf{Regression coefficients}} \\
\midrule
Fixed effect
& \textit{b}
& 95\% CI
& \textit{z}
& \textit{p} \\
\midrule
Intercept
    & 1.02 & [0.90, 1.15] & 16.14 & < .001 \\

AI-support condition: LO (NS)
    & -0.11 & [-0.29, 0.07] & -1.22 & .223 \\

AI-support condition: NIC (NS)
    & -0.22 & [-0.41, -0.04] & -2.37 & .018 \\

AI-support condition: IC (NS)
    & -0.15 & [-0.33, 0.04] & -1.58 & .114 \\
\midrule

\addlinespace[1em]
\multicolumn{5}{l}{\textbf{ANOVA omnibus tests}} \\
\midrule
\multicolumn{2}{l}{Fixed effect}
& $\chi^2$
& df
& \textit{p} \\
\midrule
\multicolumn{2}{l}{AI-support condition}
    & 5.90 & 3 & .117 \\

\bottomrule
\end{tabular}
\caption{Results of the Poisson regression on the frequency of the highest confidence rating, including AI-support condition as a fixed effect.}
\label{tab:study1-confidence6}
\end{table}

\paragraph{Study 1 -- Lowest confidence rating -- AI-support condition}
\textcolor{white}{[parkour]}

\begin{table}[!h]
\centering
\captionsetup{skip=5pt}
\begin{tabular}{lcccc}
\toprule

\multicolumn{5}{l}{\textbf{Regression coefficients}} \\
\midrule
Fixed effect
& \textit{b}
& 95\% CI
& \textit{z}
& \textit{p} \\
\midrule
Intercept
    & -1.23 & [-1.61, -0.85] & -6.28 & < .001 \\

AI-support condition: LO (NS)
    & 0.17 & [-0.34, 0.69] & 0.66 & .509 \\

AI-support condition: NIC (NS)
    & 0.17 & [-0.34, 0.69] & 0.66 & .509 \\

AI-support condition: IC (NS)
    & -0.19 & [-0.77, 0.38] & -0.65 & .515 \\
\midrule

\addlinespace[1em]
\multicolumn{5}{l}{\textbf{ANOVA omnibus tests}} \\
\midrule
\multicolumn{2}{l}{Fixed effect}
& $\chi^2$
& df
& \textit{p} \\
\midrule
\multicolumn{2}{l}{AI-support condition}
    & 2.34 & 3 & .504 \\

\bottomrule
\end{tabular}
\caption{Results of the Poisson regression on the frequency of the lowest confidence rating, including AI-support condition as a fixed effect.}
\label{tab:study1-confidence0}
\end{table}

\newpage
\textbf{Study 2 - Bird classification} In line with our expectations, the highest confidence rating in this task was relatively common, accounting for 12\% of all responses, and its frequency significantly differed across conditions (\textit{p} = .015). In particular, it was less frequent in the NS condition (\textit{M} = 1.04, \textit{SD} = 2.07) compared to the IC condition (\textit{M} = 1.52, \textit{SD} = 2.61; \textit{p} = .031), while all other contrasts involving the LO (\textit{M} = 1.29, \textit{SD} = 2.25) and the NIC conditions (\textit{M} = 1.07, \textit{SD} = 1.95) were not significant (all \textit{p} values $\geq .056$). Furthermore, the lowest confidence rating was more common in Study 2 (7\% of all responses) than in Study 1 and its frequency differed across the four conditions (\textit{p} < .001). In particular, it was significantly more frequent in the NS condition (\textit{M} = 1.01, \textit{SD} = 2.35) than in the NIC (\textit{M} = 0.37, \textit{SD} = 1.09; \textit{p} < .001) and IC (\textit{M} = 0.55, \textit{SD} = 1.34; \textit{p} = .004) conditions, while it was not different from the frequency observed in the LO condition (\textit{M} = 0.65, \textit{SD} = 1.67; \textit{p} = .058). All other contrasts were not significant (all \textit{p} vales $\geq$ .077)

\paragraph{Study 2 -- Highest confidence rating -- AI-support condition}
\textcolor{white}{[parkour]}

\begin{table}[!h]
\centering
\captionsetup{skip=5pt}
\begin{tabular}{lcccc}
\toprule

\multicolumn{5}{l}{\textbf{Regression coefficients}} \\
\midrule
Fixed effect
& \textit{b}
& 95\% CI
& \textit{z}
& \textit{p} \\
\midrule
Intercept
    & 1.04 & [0.84, 1.28] & 0.33 & .745 \\

AI-support condition: LO (NS)
    & 1.25 & [0.94, 1.65] & 1.56 & .119 \\

AI-support condition: NIC (NS)
    & 1.03 & [0.77, 1.38] & 0.22 & .826 \\

AI-support condition: IC (NS)
    & 1.47 & [1.12, 1.93] & 2.80 & .005 \\
\midrule

\addlinespace[1em]
\multicolumn{5}{l}{\textbf{ANOVA omnibus tests}} \\
\midrule
\multicolumn{2}{l}{Fixed effect}
& $\chi^2$
& df
& \textit{p} \\
\midrule
\multicolumn{2}{l}{AI-support condition}
    & 10.51 & 3 & .015 \\
\midrule

\addlinespace[1em]
\multicolumn{5}{l}{\textbf{Post-hoc contrasts}} \\
\midrule
Contrast
& \textit{b}
& \textit{SE}
& \textit{z}
& \textit{p} \\
\midrule
NS vs.\ LO
    & 0.80 & 0.11 & -1.56 & .712 \\

NS vs.\ NIC
    & 0.97 & 0.14 & -0.22 & 1 \\

NS vs.\ IC
    & 0.68 & 0.09 & -2.80 & .031 \\

LO vs.\ NIC
    & 1.21 & 0.17 & 1.35 & 1 \\

LO vs.\ IC
    & 0.85 & 0.11 & -1.26 & 1 \\

NIC vs.\ IC
    & 0.70 & 0.10 & -2.60 & .056 \\

\bottomrule
\end{tabular}
\caption{Results of the Poisson regression on the frequency of the highest confidence rating, including AI-support condition as a fixed effect.}
\label{tab:study2-confidence6}
\end{table}

\newpage
\paragraph{Study 2 -- Lowest confidence rating -- AI-support condition}
\textcolor{white}{[parkour]}

\begin{table}[!h]
\centering
\captionsetup{skip=5pt}
\begin{tabular}{lcccc}
\toprule

\multicolumn{5}{l}{\textbf{Regression coefficients}} \\
\midrule
Fixed effect
& Rate ratio
& 95\% CI
& \textit{z}
& \textit{p} \\
\midrule
Intercept
    & 1.01 & [0.82, 1.25] & 0.11 & .914 \\

AI-support condition: LO (NS)
    & 0.64 & [0.46, 0.90] & -2.59 & .010 \\

AI-support condition: NIC (NS)
    & 0.37 & [0.25, 0.55] & -4.83 & < .001 \\

AI-support condition: IC (NS)
    & 0.54 & [0.38, 0.77] & -3.40 & < .001 \\
\midrule

\addlinespace[1em]
\multicolumn{5}{l}{\textbf{ANOVA omnibus tests}} \\
\midrule
\multicolumn{2}{l}{Fixed effect}
& $\chi^2$
& df
& \textit{p} \\
\midrule
\multicolumn{2}{l}{AI-support condition}
    & 28.18 & 3 & < .001 \\
\midrule

\addlinespace[1em]
\multicolumn{5}{l}{\textbf{Post-hoc contrasts}} \\
\midrule
Contrast
& Rate ratio
& \textit{SE}
& \textit{z}
& \textit{p} \\
\midrule
NS vs.\ LO
    & 1.56 & 0.27 & 2.59 & .058 \\

NS vs.\ NIC
    & 2.72 & 0.56 & 4.83 & < .001 \\

NS vs.\ IC
    & 1.85 & 0.34 & 3.40 & .004 \\

LO vs.\ NIC
    & 1.74 & 0.39 & 2.49 & .077 \\

LO vs.\ IC
    & 1.18 & 0.24 & 0.85 & 1 \\

NIC vs.\ IC
    & 0.68 & 0.16 & -1.68 & .561 \\

\bottomrule
\end{tabular}
\caption{Results of the Poisson regression on the frequency of the lowest confidence rating, including AI-support condition as a fixed effect.}
\label{tab:study2-confidence0}
\end{table}

\newpage
\section*{Impact of inaccurate concept detection on trust}
\addcontentsline{toc}{section}{Impact of inaccurate concept detection on trust}

\bigskip
We conducted an exploratory analysis to examine whether exposure to concept detection errors decreases participants' trust in the CBM. Specifically, we focused on participants in the NIC and IC conditions, who had access to the concepts detected by the model, and assessed whether the number of incorrectly detected concepts in a trial affected participants' tendency to reject the model's prediction by selecting the opposite class (and therefore making an incorrect classification).\footnote{We focused on this behavioral measure of trust because the post-experiment trust questionnaire, administered to participants in all AI-supported conditions, did not distinguish between trust in the concept detection component and trust in the final label prediction, conflating two related but potentially distinct aspects of trust in CBMs.}

To this end, two of the authors independently annotated each of the selected stimuli from both Studies 1 and 2 for the six concepts used by the corresponding CBMs, following the concept definitions provided to participants in the user studies. Disagreements were resolved through discussion. We relied on these manual annotations rather than the concept annotations provided in \Emails and \Cub for two reasons. First, we had modified some of the original definitions and label for the features annotated in the datasets. Second, the original feature annotations in both datasets were not produced through manual human annotation. In \Emails, concept labels were generated by large language models prompted using expert annotations on a subset of emails, whereas in \Cub, concept annotations were assigned at the species level rather than individually for each image. Both procedures may have introduced inaccuracies in the concept ground truth that would have compromised the present analysis.

Initial agreement between the two annotators was 90\% for the \Cub stimuli and 74\% for the \Emails stimuli, consistent with the more subjective nature of the concepts used in Study 1. Given this difference in annotation reliability, we restricted the analysis to the \Cub dataset (Study 2).

We then fitted a logistic mixed-effects model predicting participants' classification accuracy on trials in which the CBM's label prediction was correct from the number of concept detection errors (i.e., instances in which the binarized concept activations presented to participants did not match the corresponding manually annotated ground-truth concepts), AI support condition, and their interaction. Random intercepts for participants and stimuli were included.

The interaction between AI support condition and the number of concept detection errors was significant, $\chi^2(1) = 4.12, p = .042$. Specifically, in the NIC condition, participants became less likely to follow the correct prediction provided by the CBM as the number of concept detection errors increased (OR = 0.72). In contrast, in the IC condition, concept detection errors had almost no impact on classification accuracy (OR = 1.12). A post hoc comparison of the two slopes confirmed that they differed significantly (\textit{p} = .042).

Although the observed effect was small, this exploratory analysis was likely underpowered. Furthermore, the number of concept detection errors exhibited limited variability, ranging only from 0 to 2 errors per trial, restricting the magnitude of any detectable effect. Nevertheless, the results suggest that inaccurate concept detection can lead users to reject otherwise correct model predictions, whereas allowing users to intervene on the detected concepts may mitigate this tendency.

\newpage
\paragraph{Impact of inaccurate concept detection on trust}
\textcolor{white}{[parkour]}

\begin{table}[!h]
\centering
\captionsetup{skip=5pt}
\begin{tabular}{lcccc}
\toprule

\multicolumn{5}{l}{\textbf{Regression coefficients}} \\
\midrule
Fixed effect
& Odds ratio
& 95\% CI
& \textit{z}
& \textit{p} \\
\midrule
Intercept
    & 10.64 & [3.07, 36.93] & 3.73 & < .001 \\

AI-support condition: NIC (IC)
    & 2.31 & [0.97, 5.50] & 1.90 & .058 \\

Concept detection errors
    & 1.21 & [0.56, 2.62] & 0.49 & .624 \\

NIC $\times$ Concept detection errors
    & 0.59 & [0.36, 0.98] & -2.03 & .042 \\
\midrule

\addlinespace[1em]
\multicolumn{5}{l}{\textbf{ANOVA omnibus tests}} \\
\midrule
\multicolumn{2}{l}{Fixed effect}
& $\chi^2$
& df
& \textit{p} \\
\midrule
\multicolumn{2}{l}{Intercept}
    & 13.87 & 1 & < .001 \\

\multicolumn{2}{l}{AI-support condition}
    & 3.60 & 1 & .058 \\

\multicolumn{2}{l}{Concept detection errors}
    & 0.24 & 1 & .624 \\

\multicolumn{2}{l}{AI-support condition $\times$ Concept detection errors}
    & 4.12 & 1 & .042 \\
\midrule

\addlinespace[1em]
\multicolumn{5}{l}{\textbf{Simple slopes}} \\
\midrule
AI-support condition
& Odds ratio
& 95\% CI
& \textit{z}
& \textit{p} \\
\midrule
IC
    & 1.21 & [0.56, 2.62] & 0.49 & .624 \\

NIC
    & 0.72 & [0.32, 1.59] & -0.81 & .415 \\
\bottomrule
\end{tabular}
\caption{Results of the mixed-effects logistic regression predicting participants' classification accuracy on trials in which the CBM's label prediction was correct, including AI-support condition, the number of concept detection errors in that trial, and their interaction as fixed effects.}
\label{tab:study2-concept-errors-trust}
\end{table}

\newpage

\section*{Study 1}
\addcontentsline{toc}{section}{Study 1}
\subsection*{Concepts and emails selection}
\addcontentsline{toc}{subsection}{Concepts and emails selection}
\bigskip

\paragraph{Concepts selection}

Of the 16 email-related features annotated in the \Emails dataset by \citet{toth2025phishfuzzer} that could be considered in the bottleneck of the CBM, we selected the three most frequently occurring in phishing emails ("fear" and "authority", which were collapsed into a single feature because they almost always co-occurred, "urgency", and "attachment"), which we relabeled as "problem alert", "time pressure", and "attachment interaction", respectively, to provide labels that could be more readily understandable by participants (see the Instructions section for descriptions of the concepts provided to participants). We also selected the three features most frequently occurring in legitimate emails ("curiosity", "neutral", and "reply"), relabeled as "update notification", "operational tone", and "reply request", respectively. See the CBM details section below for further information on the models' on the concepts extractor.

We note here that, due to a human error, the CBM's concept activation values were incorrect for two emails. For one email (ID 263), the incorrect values produced the same active concepts and intervention outcomes in the interface as the correct values, so we retained the corresponding observations. For the other (ID 268), they produced a different set of active concepts compared to the correct values, so we excluded the corresponding observations from the dataset. However, retaining these observations did not change the results.

\paragraph{Emails selection}
The following criteria were applied when selecting emails from the \Emails dataset introduced by \citet{toth2025phishfuzzer}.

\begin{enumerate}
\item We considered only human-written English emails fewer than 400 words. Restricting the sample to this range allowed us to obtain a relatively homogeneous set of stimuli, thereby reducing potential variability in attentional demands across items. Applying this criterion yielded a corpus of 2,330 emails. Of these, 1,064 emails were used to train the CBM, 266 were included in the validation set, and the remaining 1,000 (500 legitimate and 500 fraudulent) emails constituted the test set.

\item Among the emails included in the test set, we further selected emails aiming to maximize diversity in terms of classification difficulty.

Classification difficulty was determined based on the results of an annotation study in which 150 participants were asked to classify 20 emails each drawn from the 1,000-email test set. The procedure was identical to that experienced by participants assigned to the \textit{No support} condition in Study 1 reported in the main manuscript. After excluding participants who failed one or more attention checks and that had unfocused the experiment tab no more than five times (which brought the sample down to 134 participants), we retained only emails that presented at least two classifications from the final sample, for a total of 967 emails. These emails were then categorized as \textit{easier} if all classifications were correct, or \textit{more challenging} if at least one classification was incorrect. This categorization was used to form the final item pool so that, among both fraudulent and legitimate included, approximately half were easier and half were more challenging.

Due to the model's high prediction accuracy (92.3\%), only 74 of the 967 emails considered were misclassified: 30 fraudulent emails incorrectly predicted as legitimate and 44 legitimate emails incorrectly predicted as fraudulent. We manually inspected these emails and removed those that were near-duplicates in content or whose ground-truth labels we suspected were incorrect (e.g., emails labeled as fraudulent that appeared to be spam). This process yielded a final set of six misclassified emails, comprising three legitimate and three fraudulent examples.

To ensure a comparable level of quality among correctly classified emails, we also manually inspected correctly predicted items until we identified 12 legitimate and 12 fraudulent emails suitable for inclusion in the user study, aiming to maximize topic diversity in the selected emails. We limited this set to 24 items to avoid drawing correctly and incorrectly classified items from pools of substantially different sizes in the user study.

For completeness, we note that in the annotation study an additional sample of 150 participants completed the same classification task while receiving the label predicted by the CBM, performing a task largely identical to that experienced by participants in the \textit{Label only} condition of Study 1. In particular, the random sampling of items was again stratified so that participants interacted with a model exhibiting 80\% accuracy. These data were collected both to create a dataset for a separate project and to assess whether participants exhibited variability in performance improvements or decrements when receiving AI support compared to performing the task unaided.

\end{enumerate}

Importantly, for both the pilot study and Study 1, we anonymized all emails by removing sender addresses and replacing identifiable information, such as person and company names, with placeholders. As our aim was to evaluate the effectiveness of CBMs rather than to simulate a realistic email-classification task, we sought to minimize the availability of cues not represented among the CBM concepts. For example, a suspicious sender address is highly predictive of a phishing attempt, yet this information was not encoded among the concepts used by the CBM. Retaining such cues could therefore have allowed participants to classify emails while disregarding the information provided by the model, weakening the experimental manipulation. 

\bigskip
\subsection*{CBM details}
\addcontentsline{toc}{subsection}{CBM details}

The split is nominally produced by a stratified \texttt{train\_test\_split}
(\texttt{scikit-learn}, \texttt{test\_size=0.2}, \texttt{random\_state=42}) over the full 2{,}330-row usable set, stratified on a combined key of task label
and a word-count quantile bin (\texttt{pandas.qcut} into 6 bins), so that both class balance and message length are preserved across the resulting
train/validation partition.

\paragraph{Concept extractor} Emails are encoded using the \texttt{sentence-transformers}' \texttt{all-MiniLM-L6-v2}
model. Encoding happens once, over the entire usable set, \emph{before} any
train/validation/test split exists.

The concept extractor is one binary linear SVM per concept (six
independent classifiers, via scikit-learn's \texttt{MultiOutputClassifier}),
trained on the \emph{full} training split.
For the SVM, a linear kernel with $C=1.0$ was
selected empirically.

\begin{table}[h]
\centering
\captionsetup{skip=5pt}
\begin{tabular}{lrrrr}
\toprule
Concept & Precision & Recall & F1 & Support \\
\midrule
Fear+Authority   & 0.89 & 0.91 & 0.90 & 501 \\
Urgency          & 0.88 & 0.94 & 0.91 & 553 \\
Curiosity        & 0.62 & 0.50 & 0.55 & 262 \\
Neutral          & 0.94 & 0.82 & 0.88 & 416 \\
Reply            & 0.89 & 0.24 & 0.38 & 33 \\
Open attachment  & 0.71 & 0.07 & 0.14 & 67 \\
\midrule
Micro avg    & 0.87 & 0.80 & 0.83 & 1{,}832 \\
Macro avg    & 0.82 & 0.58 & 0.63 & 1{,}832 \\
Weighted avg & 0.86 & 0.80 & 0.81 & 1{,}832 \\
\bottomrule
\end{tabular}
\caption{Emails concept extractor: per-concept classification report on the
1{,}000 test set used for the annotation study.}
\end{table}

\begin{figure}[t]
    \centering
    \includegraphics[width=.9\linewidth]{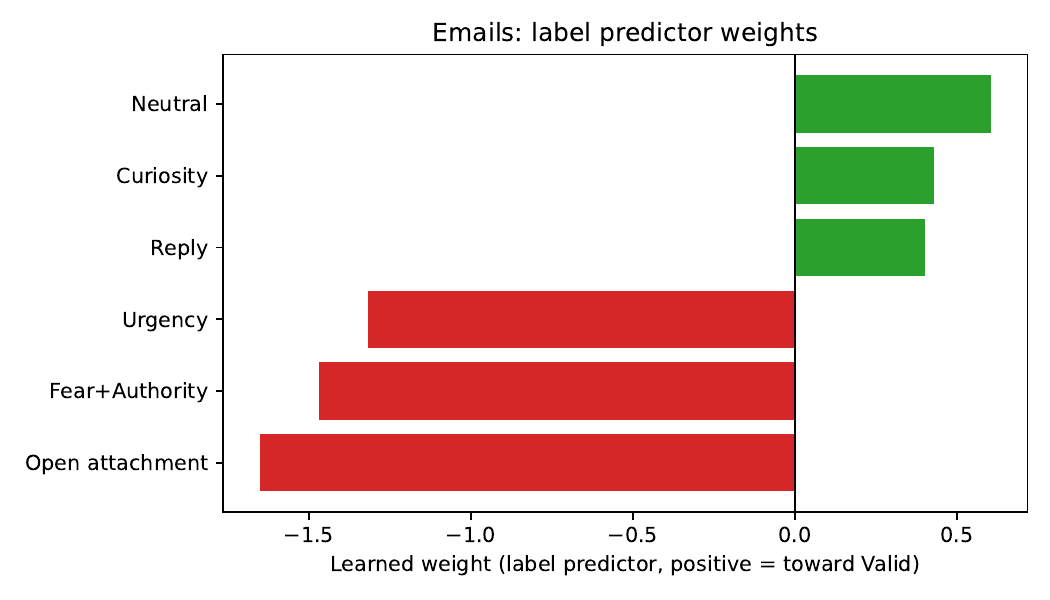}
    \caption{Logistic regression weights for the six concepts used in Study 1 (\Emails dataset). The bias term is fixed at 0.}
    \label{fig:label_predictor_weights_emails}
\end{figure}
    
\paragraph{Task predictor}
In the logistic regression parameters, we set \texttt{fit\_intercept=False}, \texttt{penalty=None} and
\texttt{class\_weight='balanced'} to compensate for the training split's class
imbalance (185 valid vs.\ 366 phishing in the training split with binary
labels).

\begin{table}[h]
\centering
\captionsetup{skip=5pt}
\begin{tabular}{lrrrr}
\toprule
Class & Precision & Recall & F1 & Support \\
\midrule
Phishing (0) & 0.93 & 0.92 & 0.93 & 500 \\
Valid (1)    & 0.92 & 0.94 & 0.93 & 500 \\
\midrule
\textbf{Accuracy} & \multicolumn{4}{r}{\textbf{0.93} (1{,}000 / 1{,}000)} \\
\bottomrule
\end{tabular}
\caption{Emails label predictor evaluated on
ground-truth concepts (bypassing the concept extractor), on the 1{,}000-row
test set. This corresponds to the upper bound on task accuracy given perfect concept
predictions.}
\end{table}

\newpage

\subsection*{Instructions}
\addcontentsline{toc}{subsection}{Instructions}

\begin{figure*}[!h]
\centering
\includegraphics[height=11.5cm]{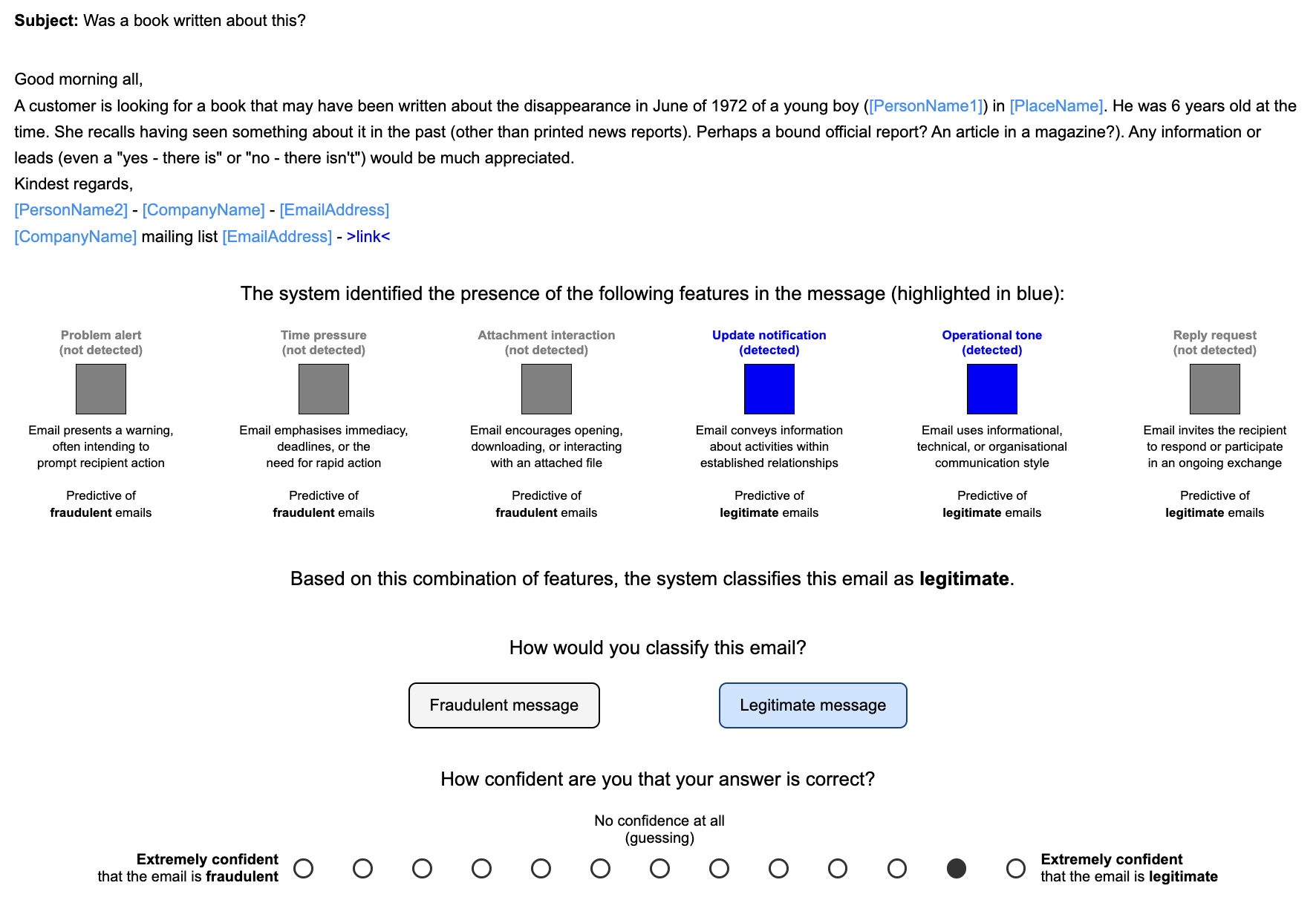}
\caption{Interface presented in the NIC and IC conditions of Study 1. The
interface presented in the LO and NS conditions were largely identical, except that the concepts information and, in the
latter case, also the model prediction, were removed.}
\label{interfaceS1}
\end{figure*}

See \cref{interfaceS1} for an example of the interface presented to participants in Study 1.

\medskip
\separator Page 1\separator

In this study, you will be shown \textbf{10 emails}, one at a time.\\
For each email, you will see the \textbf{subject line} and the \textbf{body of the message}.
\textbf{Your task is to classify each email} as either\\
\textbf{legitimate} (i.e., the sender does not intend to harm the recipient) or\\
\textbf{fraudulent} (i.e., the sender intends to harm the recipient,for example installing malware or harvesting personal information).

After having provided your answer, you will also be asked \textbf{how confident you are in your classification}.\\
Once the confidence scale appears, you will still be able to revise your classification if needed.

\textbf{No feedback} will be given on \textbf{whether your classification is correct}.

\textbf{IMPORTANT}\\
To ensure data quality, it is important that you \textbf{remain on the study page for the entire duration of the task}.\\
Participants who do not follow the instructions may be \textbf{excluded from future studies conducted by our research group}.

\medskip
\separator Page 2\separator

The emails you will be presented with are \textbf{real} and may contain a variety of messages,\\
such as alerts, requests, and summaries of blog thread exchanges.

For this reason, we \textbf{removed the sender's email address} and \textbf{anonymised sensitive information}.\\
For example, we replaced the names of individuals and companies with the tags \\``{\color{dodgerblue}[PersonName]}'' and ``{\color{dodgerblue}[CompanyName]}'', respectively.

Additionally, any \textbf{clickable text} in the original email is indicated by enclosing it within the \textbf{symbols ><}, for example: ``{\color{linkblue}>link<}''.

\medskip
\separator Page 3\separator

\noindent\hl{[LO condition]}

An \textbf{AI system}, trained to classify emails, will perform the same task as you.

Below the body of each email,\\
you will be informed \textbf{whether the system classified the email as legitimate or fraudulent}.

\noindent\hl{[NIC / IC conditions]}

An \textbf{AI system}, trained to classify emails, will perform the same task as you.

First, the system analyses the content of each email\\
to identify whether certain \textbf{features} are present.

The features the system looks for are listed below.

\begin{itemize}
    \item \textbf{Problem alert} (predictive of \textit{fraudulent} emails)\\The email presents a warning or security issue, often with the intention of prompting the recipient to take action.
    \item \textbf{Urgency cue} (predictive of \textit{fraudulent} emails)\\The email emphasises immediacy, deadlines, or the need for rapid action.
    \item \textbf{Attachment interaction} (predictive of \textit{fraudulent} emails)\\The email encourages the recipient to open, download, or otherwise interact with an attached file.
    \item \textbf{Update notification} (predictive of \textit{legitimate} emails)\\The email conveys information about activities or interests within an established service or relationship.
    \item \textbf{Operational tone} (predictive of \textit{legitimate} emails)\\The email uses an informational, procedural, technical, or organisational communication style.
    \item \textbf{Reply request} (predictive of \textit{legitimate} emails)\\The email invites the recipient to respond, provide feedback, or participate in an ongoing exchange.
\end{itemize}

These features will be presented to you, below the body of each email, as a series of boxes. 

Each box corresponds to one feature and they will be coloured\\in \textbf{blue if the system detected the corresponding feature} in the email or in \textbf{grey if it did not}.

Based on the combination of detected features, the system then classifies the email as either \textbf{legitimate} or \textbf{fraudulent}.\\
This classification is shown below the feature boxes.

The features are intended to help \textbf{explain the system's classification}.\\
In other words, they show which aspects of the email led the system to produce its classification.

Please note that the \textbf{same combination of detected features} may lead to \textbf{different classifications},\\as the system considers not only whether a feature is present, but also how strongly it detects it in a specific email.

\medskip
\separator Page 4\separator

\noindent\hl{[IC condition]}

You will also be able to \textbf{interact with the AI system}.

Specifically, you can \textbf{click on the feature boxes to change their state} (from detected to not detected, or vice versa).\\
This allows you to explore \textbf{whether the system's classification would change as well}.

For example, you may \textbf{unselect a feature} that the system has identified if you do not believe it is present in the email,\\
or, conversely, \textbf{select a feature} that the system has not identified if you believe it is present.\\
After making changes, you will see whether the system's classification updates accordingly.\\
This information may help you decide \textbf{how much to trust the system's classification}.

In some cases, you may agree with the system's classification but still disagree with the set of features it has identified.\\
If so, in addition to providing your classification, please \textbf{adjust the set of features} so that it reflects your assessment.

\medskip
\separator Page 5\separator

\noindent\hl{[LO / NIC / IC conditions]}

After classifying all the emails,\\you will be asked a few questions about \textbf{your interaction with the AI system}\\and to rate your \textbf{overall level of experience with AI systems}.

\medskip
\separator Page 6\separator

\textbf{IMPORTANT}\\
After data collection is complete, \textbf{the 5 most accurate participants will receive a bonus of \pounds5.00}.\\ Therefore, when classifying each email, please \textbf{aim to be as accurate as possible}.

\noindent\hl{[LO / NIC / IC conditions]}

When providing your classifications,\\
keep in mind that although the \textbf{AI system} has been trained for this task, it \textbf{can still make mistakes}.\\
\textbf{Blindly trusting the system} may therefore lead to errors and \textbf{negatively affect your overall accuracy}.

\clearpage
\subsection*{Tables of statistical analyses results}
\addcontentsline{toc}{subsection}{Tables of statistical analysis results}

In the following tables, all predictors were categorical and dummy coded (reference level reported in parentheses). For post-hoc comparisons, Bonferroni-corrected \textit{p} values are reported.

\paragraph{Accuracy - Only AI-support condition (all participants)}
\textcolor{white}{[parkour]}

\begin{table}[h]
\centering
\captionsetup{skip=5pt}
\begin{tabular}{lcccc}
\toprule

\multicolumn{5}{l}{\textbf{Regression coefficients}} \\
\midrule
Fixed effect
& Odds ratio
& 95\% CI
& \textit{z}
& \textit{p} \\
\midrule
Intercept
    & 3.81 & [2.44, 5.96] & 5.86 & < .001 \\
AI-support condition: LO (NS)
    & 1.44 & [1.11, 1.89] & 2.70 & .007 \\
AI-support condition: NIC (NS)
    & 1.75 & [1.33, 2.31] & 4.00 & < .001 \\
AI-support condition: IC (NS)
    & 1.65 & [1.25, 2.18] & 3.54 & < .001 \\

\midrule
\addlinespace[1em]
\multicolumn{5}{l}{\textbf{ANOVA Omnibus tests}} \\
\midrule
\multicolumn{2}{l}{Fixed effect}
& $\chi^2$
& df
& \textit{p} \\
\midrule
\multicolumn{2}{l}{AI-support condition}
    & 19.78 & 3 & < .001 \\

\midrule
\addlinespace[1em]
\multicolumn{5}{l}{\textbf{Post-hoc contrasts}} \\
\midrule
Contrast
& Odds ratio
& \textit{SE}
& \textit{z}
& \textit{p} \\
\midrule
NS vs.\ LO
    & 0.69 & 0.09 & -2.70 & .043 \\
NS vs.\ NIC
    & 0.57 & 0.08 & -4.00 & < .001 \\
NS vs.\ IC
    & 0.61 & 0.09 & -3.54 & .002 \\
LO vs.\ NIC
    & 0.82 & 0.12 & -1.36 & 1 \\
LO vs.\ IC
    & 0.87 & 0.13 & -0.94 & 1 \\
NIC vs.\ IC
    & 1.06 & 0.16 & 0.40 & 1 \\

\bottomrule
\end{tabular}
\caption{Results of the mixed-effects logistic regression on participants' accuracy, including only AI-support condition as fixed effect.}
\label{tab:study1-accuracy-simplemodel}
\end{table}

\newpage
\paragraph{Accuracy - Only AI-support condition (excluding non-interactive participants)}
\textcolor{white}{[parkour]}

\begin{table}[!h]
\centering
\captionsetup{skip=5pt}
\begin{tabular}{lcccc}
\toprule

\multicolumn{5}{l}{\textbf{Regression coefficients}} \\
\midrule
Fixed effect
& Odds ratio
& 95\% CI
& \textit{z}
& \textit{p} \\
\midrule
Intercept
    & 3.94 & [2.47, 6.28] & 5.77 & < .001 \\
AI-support condition: LO (NS)
    & 1.45 & [1.10, 1.90] & 2.67 & .008 \\
AI-support condition: NIC (NS)
    & 1.76 & [1.33, 2.32] & 3.96 & < .001 \\
AI-support condition: IC (NS)
    & 2.04 & [1.47, 2.84] & 4.22 & < .001 \\
\midrule

\addlinespace[1em]
\multicolumn{5}{l}{\textbf{ANOVA omnibus tests}} \\
\midrule
\multicolumn{2}{l}{Fixed effect}
& $\chi^2$
& df
& \textit{p}\\
\midrule
\multicolumn{2}{l}{AI-support condition}
    & 24.08 & 3 & < .001 \\
\midrule

\addlinespace[1em]
\multicolumn{5}{l}{\textbf{Post-hoc contrasts}} \\
\midrule
Contrast
& Odds ratio
& \textit{SE}
& \textit{z}
& \textit{p} \\
\midrule
NS vs.\ LO
    & 0.69 & 0.10 & -2.67 & .046 \\
NS vs.\ NIC
    & 0.57 & 0.08 & -3.96 & < .001 \\
NS vs.\ IC
    & 0.49 & 0.08 & -4.22 & < .001 \\
LO vs.\ NIC
    & 0.82 & 0.12 & -1.35 & 1 \\
LO vs.\ IC
    & 0.71 & 0.12 & -2.03 & .257 \\
NIC vs.\ IC
    & 0.86 & 0.15 & -0.87 & 1 \\

\bottomrule
\end{tabular}
\caption{Results of the mixed-effects logistic regression on participants'
accuracy, including only AI-support condition as a fixed effect.}
\label{tab:study1-accuracy-simplemodel-nononinteractive}
\end{table}

\newpage
\paragraph{Accuracy - AI-support condition, Item's ground truth, and Model's prediction accuracy (all participants)}
\textcolor{white}{[parkour]}

\begin{table}[!h]
\centering
\captionsetup{skip=5pt}

\begin{tabular}{lcccc}
\toprule

\multicolumn{5}{l}{\textbf{Regression coefficients}} \\
\midrule
Fixed effect
& Odds ratio
& 95\% CI
& \textit{z}
& \textit{p} \\
\midrule
Intercept
    & 2.64 & [0.80, 8.70] & 1.60 & .110 \\

AI-support condition: LO (NS)
    & 0.53 & [0.28, 1.00] & -1.97 & .049 \\

AI-support condition: NIC (NS)
    & 0.91 & [0.47, 1.77] & -0.27 & .786 \\

AI-support condition: IC (NS)
    & 0.99 & [0.50, 1.94] & -0.04 & .966 \\

Item ground-truth (legitimate)
    & 0.79 & [0.15, 4.29] & -0.27 & .786 \\

Model classification accuracy (wrong)
    & 1.34 & [0.35, 5.13] & 0.42 & .674 \\

LO $\times$ Item ground-truth
    & 1.14 & [0.45, 2.88] & 0.28 & .783 \\

NIC $\times$ Item ground-truth
    & 1.52 & [0.59, 3.92] & 0.87 & .385 \\

IC $\times$ Item ground-truth
    & 0.73 & [0.28, 1.92] & -0.63 & .527 \\

LO $\times$ Model classification accuracy
    & 4.02 & [1.90, 8.51] & 3.63 & $< .001$ \\

NIC $\times$ Model classification accuracy
    & 2.87 & [1.31, 6.28] & 2.64 & .008 \\

IC $\times$ Model classification accuracy
    & 3.17 & [1.41, 7.11] & 2.79 & .005 \\

Item ground-truth $\times$ Model classification accuracy
    & 1.73 & [0.26, 11.72] & 0.56 & .573 \\

LO $\times$ Item ground-truth $\times$ Model classification accuracy
    & 0.91 & [0.30, 2.75] & -0.17 & .867 \\

NIC $\times$ Item ground-truth $\times$ Model classification accuracy
    & 0.41 & [0.13, 1.25] & -1.57 & .116 \\

IC $\times$ Item ground-truth $\times$ Model classification accuracy
    & 0.67 & [0.21, 2.12] & -0.68 & .498 \\
\midrule

\addlinespace[1em]
\multicolumn{5}{l}{\textbf{ANOVA omnibus tests}} \\
\midrule
\multicolumn{2}{l}{Fixed effect}
& $\chi^2$
& df
& \textit{p}\\
\midrule
\multicolumn{2}{l}{Intercept}
    & 2.55 & 1 & .110 \\
\multicolumn{2}{l}{AI-support condition}
    & 5.59 & 3 & .133 \\
\multicolumn{2}{l}{Item ground-truth}
    & 0.07 & 1 & .786 \\
\multicolumn{2}{l}{Model classification accuracy}
    & 0.18 & 1 & .674 \\
\multicolumn{2}{l}{AI-support condition $\times$ Item ground-truth}
    & 2.35 & 3 & .504 \\
\multicolumn{2}{l}{AI-support condition $\times$ Model classification accuracy}
    & 15.04 & 3 & .002 \\
\multicolumn{2}{l}{Item ground-truth $\times$ Model classification accuracy}
    & 0.32 & 1 & .573 \\
\multicolumn{2}{l}{AI-support condition $\times$ Item ground-truth $\times$ Model classification accuracy}
    & 2.97 & 3 & .396 \\
\midrule

\addlinespace[1em]
\multicolumn{5}{l}{\textbf{Post-hoc contrasts}} \\
\midrule
Contrast
& Estimate
& \textit{SE}
& \textit{z}
& \textit{p} \\
\midrule
\multicolumn{5}{l}{\textit{Model prediction wrong}} \\
NS vs.\ LO
    & 0.57 & 0.24 & 2.36 & .218 \\
NS vs.\ NIC
    & -0.12 & 0.25 & -0.48 & 1.000 \\
NS vs.\ IC
    & 0.17 & 0.25 & 0.68 & 1.000 \\
LO vs.\ NIC
    & -0.69 & 0.24 & -2.88 & .047 \\
LO vs.\ IC
    & -0.40 & 0.24 & -1.66 & 1.000 \\
NIC vs.\ IC
    & 0.29 & 0.25 & 1.16 & 1.000 \\

\addlinespace[0.5em]
\multicolumn{5}{l}{\textit{Model prediction correct}} \\
NS vs.\ LO
    & -0.77 & 0.16 & -4.74 & $< .001$ \\
NS vs.\ NIC
    & -0.72 & 0.16 & -4.42 & $< .001$ \\
NS vs.\ IC
    & -0.78 & 0.17 & -4.64 & $< .001$ \\
LO vs.\ NIC
    & 0.05 & 0.17 & 0.27 & 1.000 \\
LO vs.\ IC
    & -0.01 & 0.18 & -0.08 & 1.000 \\
NIC vs.\ IC
    & -0.06 & 0.18 & -0.34 & 1.000 \\

\bottomrule
\end{tabular}
\caption{Results of the mixed-effects logistic regression on participants' accuracy, including AI-support condition, correct answer, model-answer correctness, and their interactions as fixed effects.}
\label{tab:study1-accuracy-fullmodel}
\end{table}

\newpage
\paragraph{Accuracy -- AI-support condition, item's ground truth, and model prediction accuracy (excluding non-interactive participants)}
\textcolor{white}{[parkour]}

\begin{table}[!h]
\centering
\captionsetup{skip=5pt}
\begin{tabular}{lcccc}
\toprule

\multicolumn{5}{l}{\textbf{Regression coefficients}} \\
\midrule
Fixed effect
& Odds ratio
& 95\% CI
& \textit{z}
& \textit{p} \\
\midrule
Intercept
    & 2.66 & [0.77, 9.18] & 1.55 & .122 \\

AI-support condition: LO (NS)
    & 0.53 & [0.28, 1.00] & -1.97 & .049 \\

AI-support condition: NIC (NS)
    & 0.91 & [0.47, 1.77] & -0.27 & .784 \\

AI-support condition: IC (NS)
    & 1.86 & [0.80, 4.35] & 1.44 & .149 \\

Item ground truth (legitimate)
    & 0.78 & [0.13, 4.52] & -0.28 & .781 \\

Model prediction accuracy (correct)
    & 1.35 & [0.33, 5.45] & 0.42 & .678 \\

LO $\times$ Item ground truth
    & 1.15 & [0.45, 2.90] & 0.29 & .774 \\

NIC $\times$ Item ground truth
    & 1.53 & [0.59, 3.94] & 0.88 & .382 \\

IC $\times$ Item ground truth
    & 0.36 & [0.11, 1.11] & -1.78 & .076 \\

LO $\times$ Model prediction accuracy
    & 4.05 & [1.91, 8.59] & 3.65 & < .001 \\

NIC $\times$ Model prediction accuracy
    & 2.88 & [1.31, 6.31] & 2.64 & .008 \\

IC $\times$ Model prediction accuracy
    & 1.87 & [0.69, 5.07] & 1.22 & .222 \\

Item ground truth $\times$ Model prediction accuracy
    & 1.86 & [0.25, 13.55] & 0.61 & .541 \\

LO $\times$ Item ground truth $\times$ Model prediction accuracy
    & 0.91 & [0.30, 2.75] & -0.17 & .866 \\

NIC $\times$ Item ground truth $\times$ Model prediction accuracy
    & 0.41 & [0.13, 1.25] & -1.57 & .116 \\

IC $\times$ Item ground truth $\times$ Model prediction accuracy
    & 1.66 & [0.42, 6.51] & 0.73 & .466 \\
\midrule

\addlinespace[1em]
\multicolumn{5}{l}{\textbf{ANOVA omnibus tests}} \\
\midrule
\multicolumn{2}{l}{Fixed effect}
& $\chi^2$
& df
& \textit{p}\\
\midrule
\multicolumn{2}{l}{Intercept}
    & 2.39 & 1 & .122 \\

\multicolumn{2}{l}{AI-support condition}
    & 10.29 & 3 & .016 \\

\multicolumn{2}{l}{Item ground truth}
    & 0.08 & 1 & .781 \\

\multicolumn{2}{l}{Model prediction accuracy}
    & 0.17 & 1 & .678 \\

\multicolumn{2}{l}{AI-support condition $\times$ Item ground truth}
    & 6.66 & 3 & .084 \\

\multicolumn{2}{l}{AI-support condition $\times$ Model prediction accuracy}
    & 14.41 & 3 & .002 \\

\multicolumn{2}{l}{Item ground truth $\times$ Model prediction accuracy}
    & 0.37 & 1 & .541 \\

\multicolumn{2}{l}{AI-support condition $\times$ Item ground truth $\times$ Model prediction accuracy}
    & 4.76 & 3 & .191 \\
\midrule

\addlinespace[1em]
\multicolumn{5}{l}{\textbf{Post-hoc contrasts}} \\
\midrule
Contrast
& Estimate
& \textit{SE}
& \textit{z}
& \textit{p} \\
\midrule
\multicolumn{5}{l}{\textit{Model prediction wrong}} \\
NS vs.\ LO
    & 0.58 & 0.24 & 2.36 & .220 \\

NS vs.\ NIC
    & -0.12 & 0.25 & -0.48 & 1 \\

NS vs.\ IC
    & -0.11 & 0.30 & -0.36 & 1 \\

LO vs.\ NIC
    & -0.69 & 0.24 & -2.88 & .048 \\

LO vs.\ IC
    & -0.68 & 0.29 & -2.34 & .229 \\

NIC vs.\ IC
    & 0.01 & 0.30 & 0.04 & 1 \\

\addlinespace[0.5em]
\multicolumn{5}{l}{\textit{Model prediction correct}} \\
NS vs.\ LO
    & -0.78 & 0.16 & -4.72 & < .001 \\

NS vs.\ NIC
    & -0.73 & 0.17 & -4.39 & < .001 \\

NS vs.\ IC
    & -0.98 & 0.21 & -4.81 & < .001 \\

LO vs.\ NIC
    & 0.05 & 0.18 & 0.28 & 1 \\

LO vs.\ IC
    & -0.21 & 0.21 & -0.99 & 1 \\

NIC vs.\ IC
    & -0.26 & 0.21 & -1.21 & 1 \\

\bottomrule
\end{tabular}
\caption{Results of the mixed-effects logistic regression on participants'
accuracy, including AI-support condition, item ground truth, model prediction
accuracy, and their interactions as fixed effects.}
\label{tab:study1-accuracy-fullmodel-nononinteractive}
\end{table}

\newpage
\paragraph{Intervention behavior - Model's concept detection accuracy (all participants)}
\textcolor{white}{[parkour]}

\begin{table}[!h]
\centering
\captionsetup{skip=5pt}
\begin{tabular}{lcccc}
\toprule

\multicolumn{5}{l}{\textbf{Regression coefficients}} \\
\midrule
Fixed effect
& Odds ratio
& 95\% CI
& \textit{z}
& \textit{p} \\
\midrule
Intercept
    & 0.14 & [0.07, 0.29] & -5.35 & < .001 \\

Model's concept detection accuracy: correct (wrong)
    & 0.19 & [0.14, 0.28] & -9.03 & < .001 \\

\bottomrule
\end{tabular}
\caption{Results of the mixed-effects logistic regression on participants' interaction behavior (i.e., whether they intervened at least once on a concept in a given trial), including model's concept detection accuracy as fixed effect.}
\label{tab:study1-intervention-modeldetection}
\end{table}

\begin{figure*}[!h]
\centering
\includegraphics[height=8.5cm]{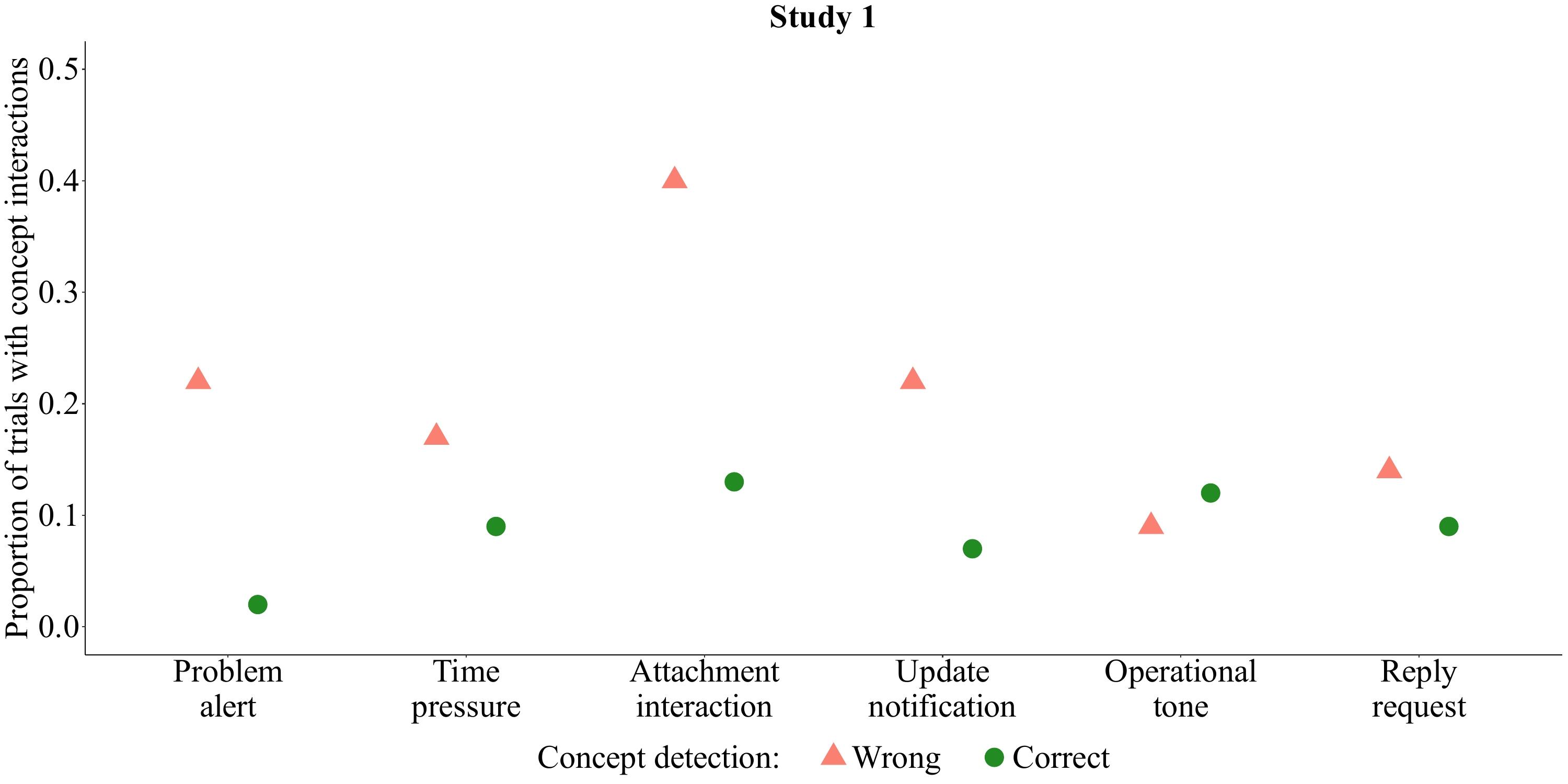}
\caption{Proportion of trials in which participants in the IC condition intervened on concept values, by model's concept detection accuracy (all participants included).}
\label{InteractionStudy1}
\end{figure*}

\paragraph{Intervention behavior - Model's concept detection accuracy (excluding non-interactive participants)}
\textcolor{white}{[parkour]}

\begin{table}[!h]
\centering
\captionsetup{skip=5pt}
\begin{tabular}{lcccc}
\toprule

\multicolumn{5}{l}{\textbf{Regression coefficients}} \\
\midrule
Fixed effect
& Odds ratio
& 95\% CI
& \textit{z}
& \textit{p} \\
\midrule
Intercept
    & 0.54 & [0.30, 0.99] & -2.01 & .045 \\

Model's concept detection accuracy: correct (wrong)
    & 0.20 & [0.14, 0.28] & -8.84 & < .001 \\

\bottomrule
\end{tabular}
\caption{Results of the mixed-effects logistic regression on participants' interaction behavior (i.e., whether they intervened at least once on a concept in a given trial), including model's concept detection accuracy as fixed effect.}
\label{tab:study1-intervention-modeldetection-nononinteractive}
\end{table}

\newpage
\paragraph{Intervention behavior - Impact of intervening on concept values on classification accuracy (all participants)}
\textcolor{white}{[parkour]}

\begin{table}[!h]
\centering
\captionsetup{skip=5pt}
\begin{tabular}{lcccc}
\toprule

\multicolumn{5}{l}{\textbf{Regression coefficients}} \\
\midrule
Fixed effect
& Odds ratio
& 95\% CI
& \textit{z}
& \textit{p} \\
\midrule
Intercept
    & 5.77 & [3.43, 9.69] & 6.62 & < .001 \\

Interaction within trial: false (true)
    & 1.28 & [0.84, 1.95] & 1.16 & .248 \\

\bottomrule
\end{tabular}
\caption{Results of the mixed-effects logistic regression on participants' accuracy, including interaction behavior (i.e., having interacted at least once within a certain trial) as fixed effect.}
\label{tab:study1-intervention-accuracy}
\end{table}

\paragraph{Intervention behavior - Impact of intervening on concept values on classification accuracy (excluding non-interactive participants)}
\textcolor{white}{[parkour]}

\begin{table}[!h]
\centering
\captionsetup{skip=5pt}
\begin{tabular}{lcccc}
\toprule

\multicolumn{5}{l}{\textbf{Regression coefficients}} \\
\midrule
Fixed effect
& Odds ratio
& 95\% CI
& \textit{z}
& \textit{p} \\
\midrule
Intercept
    & 10.26 & [4.83, 21.78] & 6.06 & < .001 \\

Interaction within trial: false (true)
    & 0.82 & [0.45, 1.48] & -0.66 & .511 \\

\bottomrule
\end{tabular}
\caption{Results of the mixed-effects logistic regression on participants' accuracy, including interaction behavior (i.e., having interacted at least once within a certain trial) as fixed effect.}
\label{tab:study1-intervention-accuracy-nononinteractive}
\end{table}

\newpage
\paragraph{Confidence - AI-support condition (all participants)}
\textcolor{white}{[parkour]}

\begin{table}[!h]
\centering
\captionsetup{skip=5pt}
\begin{tabular}{lcccc}
\toprule

\multicolumn{5}{l}{\textbf{Regression coefficients}} \\
\midrule
Fixed effect
& \textit{b}
& \textit{SE}
& \textit{z}
& \textit{p} \\
\midrule
AI-support condition: LO (NS)
    & -0.06 & 0.21 & -0.27 & .788 \\

AI-support condition: NIC (NS)
    & -0.26 & 0.21 & -1.21 & .226 \\

AI-support condition: IC (NS)
    & -0.01 & 0.21 & -0.06 & .955 \\

\midrule

\addlinespace[1em]
\multicolumn{5}{l}{\textbf{ANOVA omnibus tests}} \\
\midrule
\multicolumn{2}{l}{Fixed effect}
& $\chi^2$
& df
& \textit{p}\\
\midrule
\multicolumn{2}{l}{AI-support condition}
    & 1.93 & 3 & .588 \\
\bottomrule
\end{tabular}
\caption{Results of the mixed-effects ordinal regression on participants' confidence ratings, including AI-support condition as fixed effect.}
\label{tab:study1-intervention}
\end{table}

\begin{figure*}[!h]
\centering
\includegraphics[height=7cm]{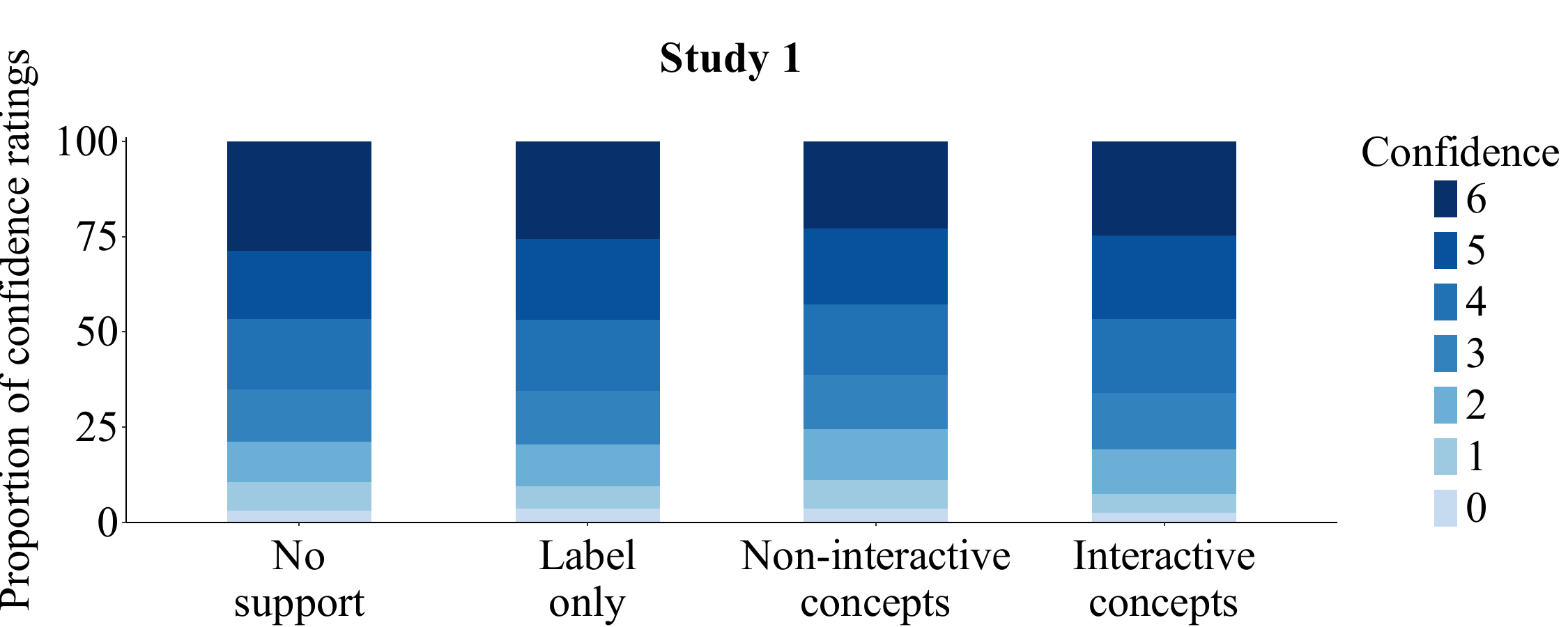}
\caption{Confidence ratings (0 = no confidence at all that the classification is correct; 6 = extreme confidence that the classification is correct) by AI-support condition (all participants included).}
\label{ConfidenceStudy1}
\end{figure*}

\paragraph{Confidence -- AI-support condition (excluding non-interactive participants)}
\textcolor{white}{[parkour]}

\begin{table}[!h]
\centering
\captionsetup{skip=5pt}
\begin{tabular}{lcccc}
\toprule

\multicolumn{5}{l}{\textbf{Regression coefficients}} \\
\midrule
Fixed effect
& \textit{b}
& \textit{SE}
& \textit{z}
& \textit{p} \\
\midrule
AI-support condition: LO (NS)
    & -0.06 & 0.21 & -0.27 & .791 \\

AI-support condition: NIC (NS)
    & -0.26 & 0.21 & -1.22 & .224 \\

AI-support condition: IC (NS)
    & -0.002 & 0.24 & -0.01 & .992 \\

\addlinespace[1em]
\multicolumn{5}{l}{\textbf{ANOVA omnibus tests}} \\
\midrule
\multicolumn{2}{l}{Fixed effect}
& $\chi^2$
& df
& \textit{p} \\
\midrule
\multicolumn{2}{l}{AI-support condition}
    & 1.93 & 3 & .587 \\
\bottomrule
\end{tabular}
\caption{Results of the mixed-effects ordinal regression on participants' confidence ratings, including AI-support condition as a fixed effect.}
\label{tab:study1-intervention-nononinteractive}
\end{table}

\newpage
\paragraph{Self-reported trust (all participants)}
\textcolor{white}{[parkour]}

\begin{table}[!h]
\centering
\captionsetup{skip=5pt}
\begin{tabular}{lccc}
\toprule

\multicolumn{4}{l}{\textbf{Kruskal--Wallis tests}} \\
\midrule
{}
& $\chi^2$
& df
& \textit{p} \\
\midrule
Trust index
    & 3.36 & 2 & .187 \\
Item 1
    & 2.25 & 2 & 1 \\
Item 2
    & 0.01 & 2 & 1 \\
Item 3
    & 1.22 & 2 & 1 \\
Item 4
    & 1.23 & 2 & 1 \\
Item 5
    & 2.45 & 2 & 1 \\
Item 6
    & 2.93 & 2 & 1 \\
Item 7
    & 5.47 & 2 & .519 \\
Item 8
    & 4.20 & 2 & .980 \\
\bottomrule
\end{tabular}
\caption{Results of the Kruskal-Wallis tests comparing the three AI-support conditions on the individual trust questionnaire items. The \textit{p} values were Bonferroni-adjusted for the eight items.}
\label{tab:study1-trust-items}
\end{table}

\paragraph{Self-reported trust (excluding non-interactive participants)}
\textcolor{white}{[parkour]}

\begin{table}[!h]
\centering
\captionsetup{skip=5pt}
\begin{tabular}{lccc}
\toprule

\multicolumn{4}{l}{\textbf{Kruskal--Wallis tests}} \\
\midrule
{}
& $\chi^2$
& df
& \textit{p} \\
\midrule
Trust index
    & 4.04 & 2 & .133 \\
Item 1
    & 3.09 & 2 & 1 \\
Item 2
    & 0.02 & 2 & 1 \\
Item 3
    & 0.70 & 2 & 1 \\
Item 4
    & 1.89 & 2 & 1 \\
Item 5
    & 1.56 & 2 & 1 \\
Item 6
    & 1.21 & 2 & 1 \\
Item 7
    & 10.38 & 2 & .044 \\
Item 8
    & 3.96 & 2 & 1 \\

\addlinespace[1em]
\multicolumn{4}{l}{\textbf{Pairwise comparisons for Item 7}} \\
\midrule
Contrast
& {}
& {}
& \textit{p} \\
\midrule
LO vs.\ NIC
    & & & .074 \\
LO vs.\ IC
    & & & .007 \\
NIC vs.\ IC
    & & & .872 \\

\bottomrule
\end{tabular}
\caption{Results of the Kruskal--Wallis tests comparing the three AI-support conditions on the trust index and individual trust questionnaire items after excluding non-interactive participants. The \textit{p} values for the individual items were Bonferroni-adjusted for eight tests. Pairwise Wilcoxon rank-sum tests for Item 7 were Bonferroni-adjusted for three comparisons.}
\label{tab:study1-trust-items-nononinteractive}
\end{table}

\clearpage
\section*{Study 2}
\addcontentsline{toc}{section}{Study 2}
\bigskip
\subsection*{Concepts and images selection}
\addcontentsline{toc}{subsection}{Concepts and images selection}

\paragraph{Concepts selection}

In \Cub, concept annotations are assigned at the species level rather than manually verified for each individual image. Consequently, to select the six concepts forming the bottleneck of the CBM, we first restricted the 112 available concepts to those whose annotations differed between the two species used in Study 2 (Le Conte's sparrow and Savannah sparrow). Among these, we excluded concepts that could be open to more subjective interpretation, such as those referring to the predominant color of the bird's body.

This process resulted in the selection of six concepts: "yellow nape", "yellow breast", and "solid belly", which were predictive of Le Conte's sparrow, and "brown crown", "white throat", and "striped breast", which were predictive of the Savannah sparrow. To improve their interpretability, we slightly revised the concept labels presented to participants, resulting in "yellow eyebrow", "yellow chest", "plain sides", "crested head", "white throat", and "striped chest", respectively (see the Instructions section for the definitions presented to participants).

\paragraph{Images selection}

Because concept annotations in \Cub are assigned at the species level, some images are annotated as containing concepts that are not actually visible (e.g., because the corresponding body part is occluded). To avoid presenting participants with concept predictions that could not be visually verified, we retained only images from the standard \Cub test set in which all body regions corresponding to the selected concepts were clearly visible. For example, we excluded images in which the bird's chest or sides were occluded or in which the bird was facing away from the camera.

After selecting the images, we conducted a pilot study on Prolific (final \textit{N} = 14 after attention-check exclusions) to assess the difficulty of the task for participants in the \textit{No Support} and \textit{Label Only} conditions. Consistent with our design objectives, the task proved challenging but feasible: participants achieved a classification accuracy of 73\% without AI support and 82\% when provided only with the CBM's predicted label.

\bigskip
\subsection*{CBM details}
\addcontentsline{toc}{subsection}{CBM details}
\paragraph{Concept extractor}
Images are encoded with CLIP (\texttt{ViT-L/14}), using CLIP's own standard
preprocessing transform.

Before any model training, each
split's embeddings are independently standardized to zero mean and unit
variance, \emph{per split}

An RBF kernel (rather than the linear
kernel used for emails) was chosen
because CLIP image embeddings and the visual concepts being predicted
(e.g.\ ``striped breast'') are not expected to be linearly separable in
embedding space the way short-text sentence embeddings are for the
email-specific concepts; \texttt{C=1.0} and \texttt{class\_weight='balanced'}

\begin{table}[h]
\centering
\captionsetup{skip=5pt}
\begin{tabular}{lrrrr}
\toprule
Concept & Precision & Recall & F1 & Support \\
\midrule
striped breast & 0.54 & 0.97 & 0.69 & 30 \\
buff breast    & 0.67 & 0.97 & 0.79 & 29 \\
white throat   & 0.82 & 0.77 & 0.79 & 30 \\
buff nape      & 0.41 & 0.59 & 0.49 & 29 \\
solid belly    & 0.93 & 0.45 & 0.60 & 29 \\
brown crown    & 0.50 & 0.30 & 0.38 & 30 \\
\midrule
Micro avg    & 0.60 & 0.67 & 0.64 & 177 \\
Macro avg    & 0.64 & 0.67 & 0.62 & 177 \\
Weighted avg & 0.64 & 0.67 & 0.62 & 177 \\
\bottomrule
\end{tabular}
\caption{CUB concept extractor: per-concept classification report on the
sparrow pair's 59 held-out test images.}
\end{table}

\paragraph{Task predictor}
Training the logistic regression with parameters \texttt{fit\_intercept=False}, \texttt{penalty='l2'} and $C=1.0$, the fitted
model reaches $100\%$ accuracy.

\begin{table}[h]
\centering
\captionsetup{skip=5pt}
\begin{tabular}{lrrrr}
\toprule
Class & Precision & Recall & F1 & Support \\
\midrule
Le Conte's & 1.00 & 1.00 & 1.00 & 29 \\
Savannah  & 1.00 & 1.00 & 1.00 & 30 \\
\midrule
\textbf{Accuracy} & \multicolumn{4}{r}{\textbf{1.00} (59 / 59)} \\
\bottomrule
\end{tabular}
\caption{CUB label predictor evaluated on
ground-truth concepts (bypassing the concept extractor), on the sparrow
pair's 59 test images. It corresponds to the upper bound on task accuracy given perfect concept predictions.}
\label{tab:task-predictor-accuracy}
\end{table}

\begin{figure}[!h]
    \centering
    \includegraphics[width=.9\linewidth]{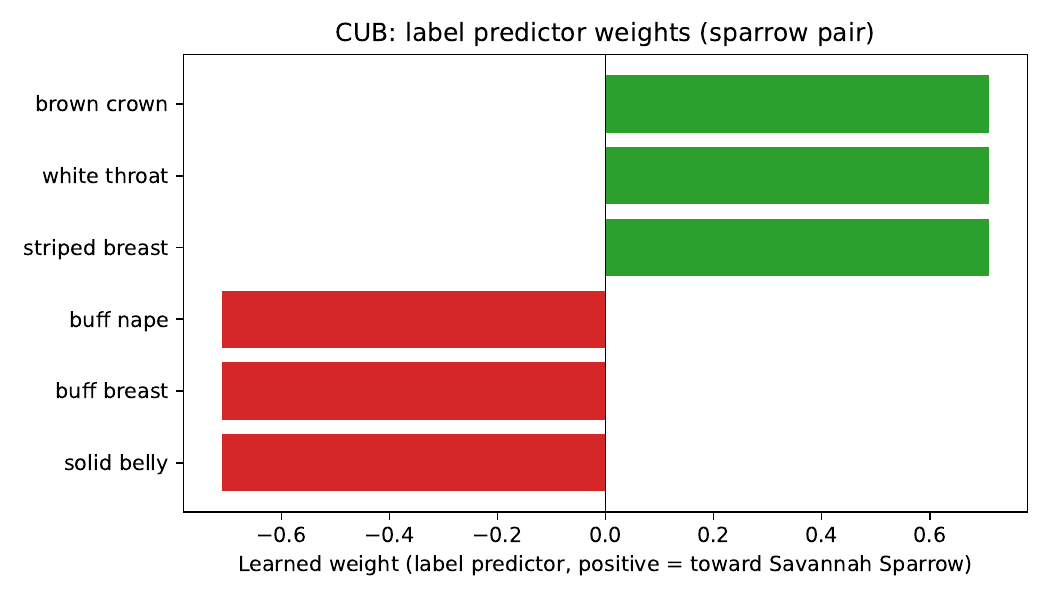}
    \caption{Logistic regression weights for the six concepts used in Study 2 (\Cub dataset). The bias term is 0.}
    \label{fig:label_predictor_weights_cub}
\end{figure}

\newpage

\subsection*{Instructions}
\addcontentsline{toc}{subsection}{Instructions}

\medskip
\separator Page 1\separator

In this study, you will be shown \textbf{10 images of birds}, one at a time.
\textbf{Your task is to classify each bird} as either a \textbf{Le Conte sparrow} or a \textbf{Savannah sparrow}.

After having provided your answer, you will also be asked \textbf{how confident you are in your classification}.\\
Once the confidence scale appears, you will still be able to revise your classification if needed.

\textbf{No feedback} will be given on \textbf{whether your classification is correct}.

\textbf{IMPORTANT}\\
To ensure data quality, it is important that you \textbf{remain on the study page for the entire duration of the task}.\\
Participants who do not follow the instructions may be \textbf{excluded from future studies conducted by our research group}.

\newpage
\separator Page 2\separator

Below, you can find three example images of each species.

Please take your time to familiarise yourself with the examples.

\medskip

\begin{center}
\setlength{\tabcolsep}{2pt}
\begin{tabular}{ccc@{\hspace{2em}}ccc}
\multicolumn{3}{c}{\textbf{Le Conte sparrow}} &
\multicolumn{3}{c}{\textbf{Savannah sparrow}} \\[0.5em]

\includegraphics[width=.135\linewidth]{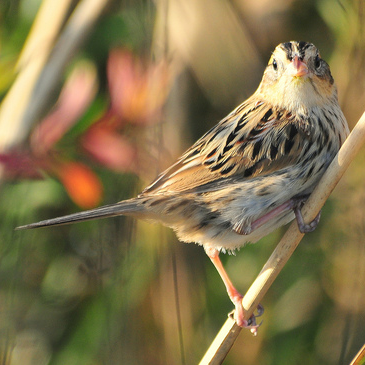} &
\includegraphics[width=.135\linewidth]{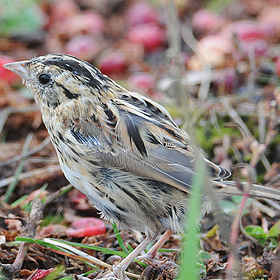} &
\includegraphics[width=.135\linewidth]{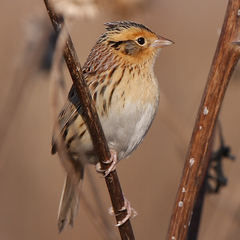} &
\includegraphics[width=.135\linewidth]{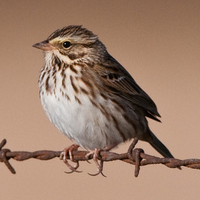} &
\includegraphics[width=.135\linewidth]{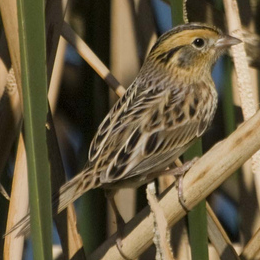} &
\includegraphics[width=.135\linewidth]{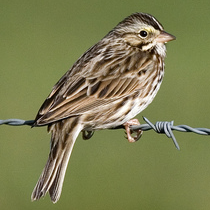}
\end{tabular}
\end{center}

\medskip
\separator Page 3\separator

\noindent\hl{[LO condition]}

An \textbf{AI system}, trained to classify bird images, will perform the same task as you.

Below each image,\\
you will be informed \textbf{whether the system classified the bird as a Le Conte sparrow or a Savannah sparrow}.

\medskip

\noindent\hl{[NIC / IC conditions]}

An \textbf{AI system}, trained to classify bird images, will perform the same task as you.

First, the system analyses each image\\
to identify whether certain \textbf{features} are present.

The features the system looks for are listed below\\ (the accompanying images illustrate the area of the bird to which each feature refers).

\begin{itemize}
    \item \textbf{Warm-coloured eyebrow} (predictive of \textit{Le Conte sparrow}):\\The predominant colour of the bird's eyebrow ranges from yellow to orange.

    \begin{center}
    \includegraphics[width=.20\linewidth]{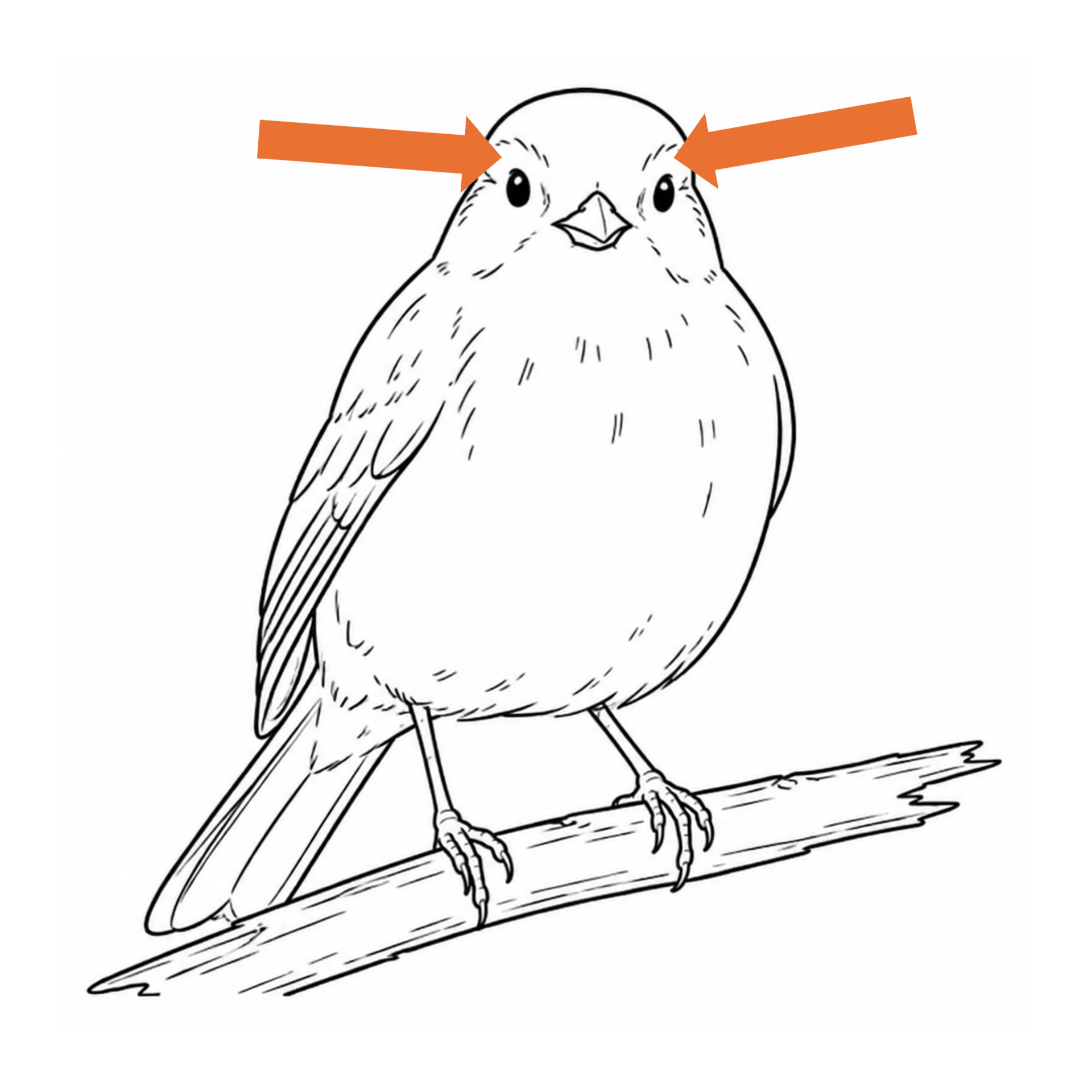}
    \end{center}

    \item \textbf{Warm-coloured chest} (predictive of \textit{Le Conte sparrow}):\\The predominant colour of the bird's chest ranges from yellow to orange.

    \begin{center}
    \includegraphics[width=.20\linewidth]{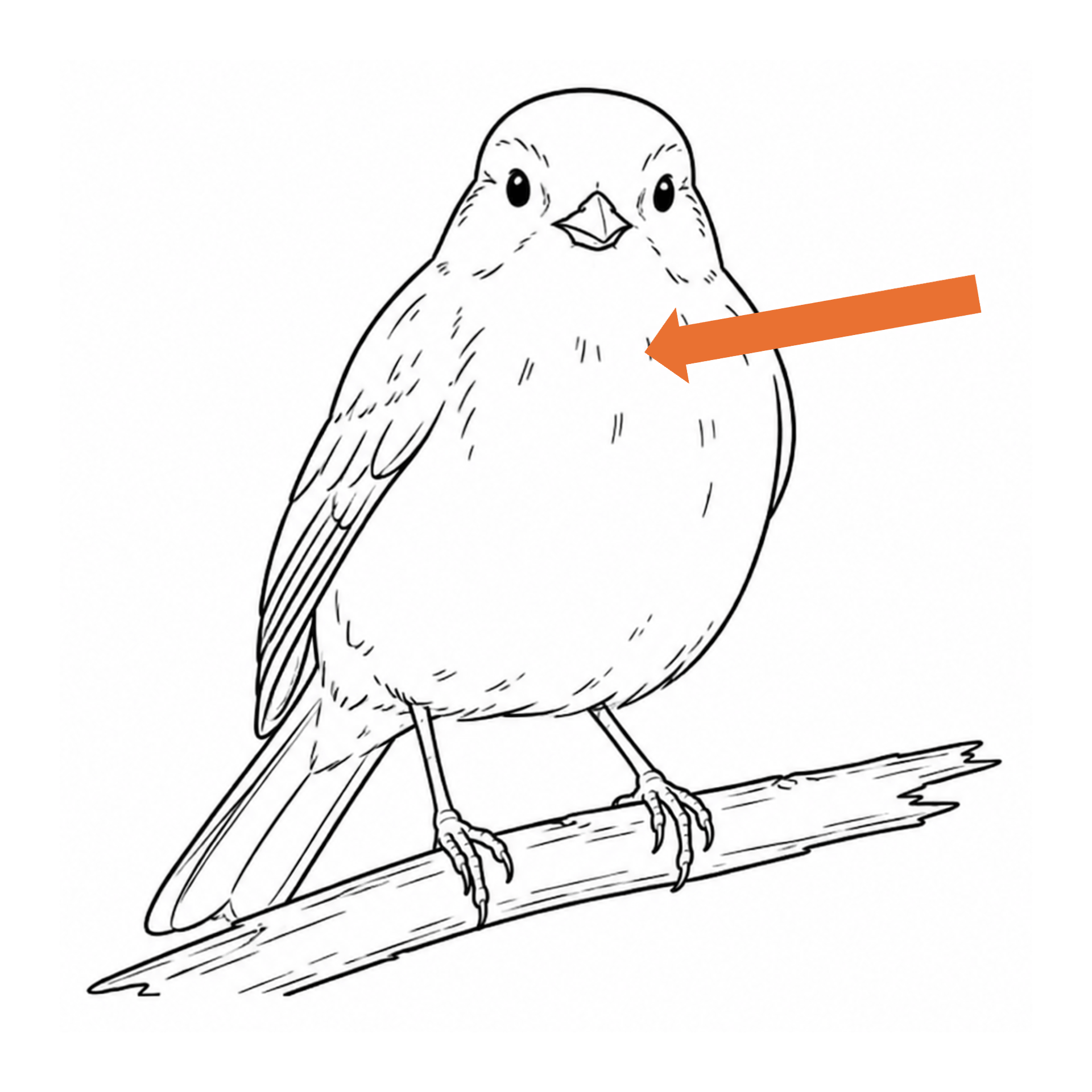}
    \end{center}

    \item \textbf{Plain sides} (predictive of \textit{Le Conte sparrow}):\\The sides of the bird's body around the belly appear plain and do not show visible patterns such as stripes, streaks, or spots.

    \begin{center}
    \includegraphics[width=.20\linewidth]{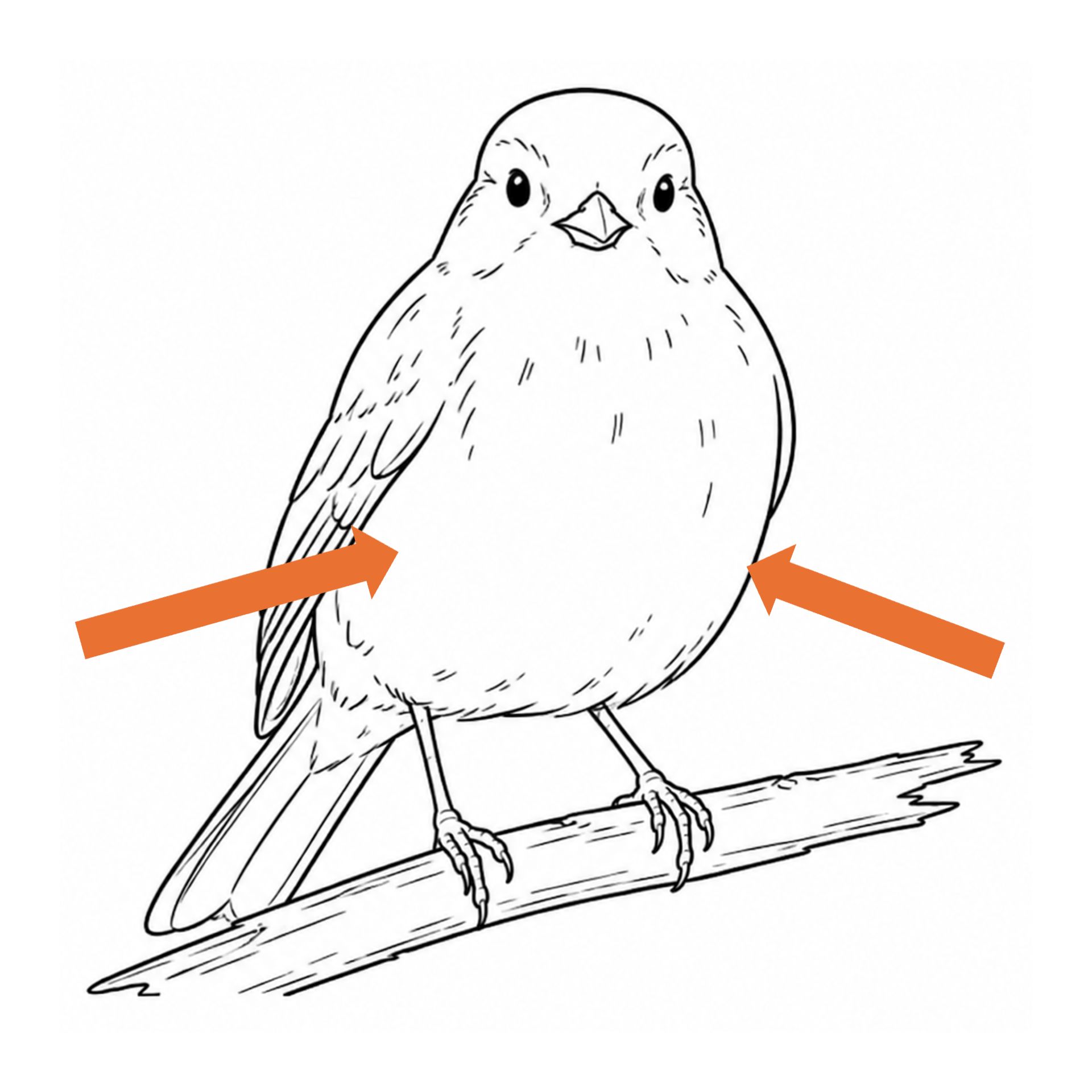}
    \end{center}

    \item \textbf{Crested head} (predictive of \textit{Savannah sparrow}):\\The bird's head presents a crest, that is, a small tuft or raised group of feathers.

    \begin{center}
    \includegraphics[width=.20\linewidth]{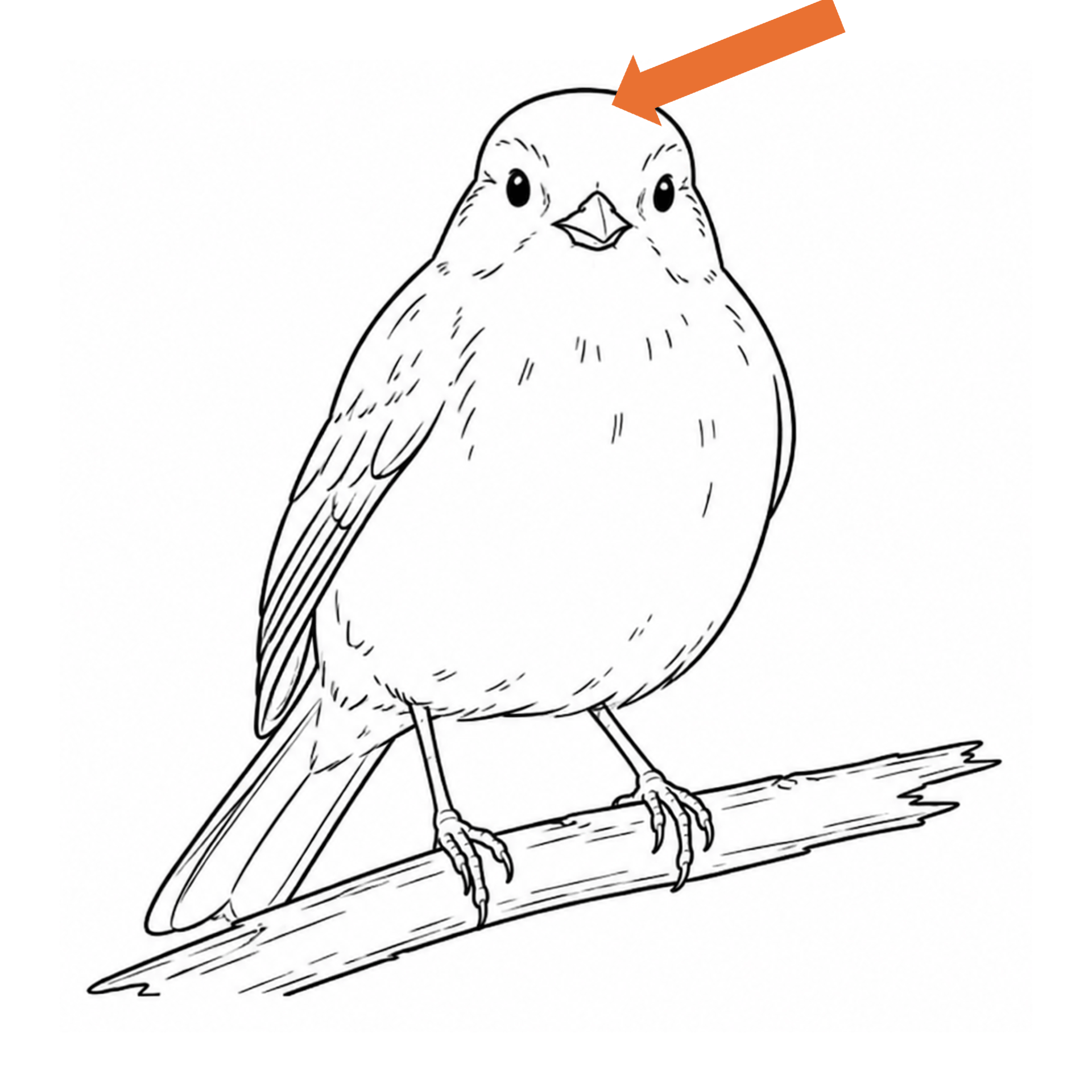}
    \end{center}

    \item \textbf{White throat} (predictive of \textit{Savannah sparrow}):\\The predominant colour of the bird's throat is white.

    \begin{center}
    \includegraphics[width=.20\linewidth]{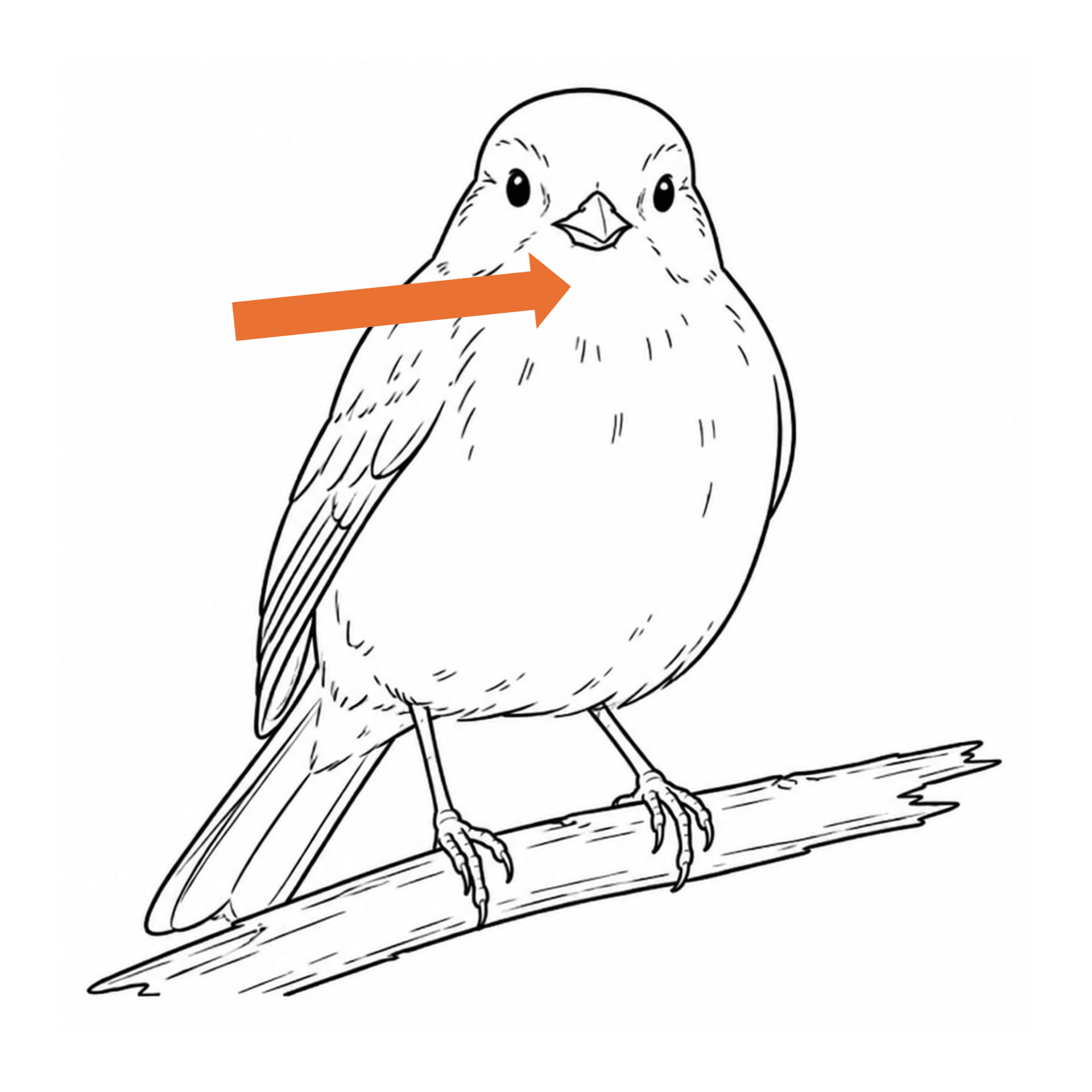}
    \end{center}

    \item \textbf{Striped chest} (predictive of \textit{Savannah sparrow}):\\The bird's chest shows visible dark stripes or streaks.

    \begin{center}
    \includegraphics[width=.20\linewidth]{Chest.png}
    \end{center}
\end{itemize}

These features will be presented to you, below each image, as a series of boxes. 

Each box corresponds to one feature:\\
it will be \textbf{coloured in blue} if the system \textbf{detected} the \textbf{feature} in the image or in \textbf{grey} if it \textbf{did not}.

Based on the combination of detected features, the system then classifies the bird\\
as either a \textbf{Le Conte sparrow} or a \textbf{Savannah sparrow}.\\
This classification is shown below the feature boxes.

The features are intended to help \textbf{explain the system's classification}.\\
In other words, they show which aspects of the image led the system to produce its classification.

Please note that the \textbf{same combination of detected features} may lead to \textbf{different classifications},\\
as the system considers not only \textbf{whether a feature is detected}, but also \textbf{how strongly it detects it} in a specific image.

\medskip
\separator Page 4\separator

\noindent\hl{[IC condition]}

You will also be able to \textbf{interact with the AI system}.

Specifically, you can \textbf{click on the feature boxes to change their state} from detected to not detected, or vice versa.\\
This allows you to explore \textbf{whether the system's classification would change as well}.

For example, you may \textbf{unselect a feature} that the system has identified if you do not believe it is present in the image\\
or, conversely, \textbf{select a feature} that the system has not identified if you believe it is present.\\
After you select or deselect any feature, the system's classification may update accordingly.\\
This information may help you decide \textbf{how much to trust the system's classification}.

In some cases, you may agree with the system's classification but still disagree with the set of features it has identified.\\
If so, in addition to providing your classification, please \textbf{adjust the set of features} so that it reflects your assessment.

\medskip
\separator Page 5\separator

\noindent\hl{[LO / NIC / IC conditions]}

After classifying all the bird images,\\
you will be asked a few questions about \textbf{your interaction with the AI system}\\
and to rate your \textbf{overall level of experience with AI systems}.

\medskip
\separator Page 6\separator

\textbf{IMPORTANT}\\
After data collection is complete, \textbf{the 5 most accurate participants will receive a bonus of \pounds5.00}.\\
Therefore, when classifying each bird, please \textbf{aim to be as accurate as possible}.

\medskip

\noindent\hl{[LO / NIC / IC conditions]}

When providing your classifications,\\keep in mind that although the \textbf{AI system} has been trained for this task, it \textbf{can still make mistakes}.\\
\textbf{Blindly trusting the system} may therefore lead to errors and \textbf{negatively affect your overall accuracy}.

\newpage
\subsection*{Tables of statistical analyses results}
\addcontentsline{toc}{subsection}{Tables of statistical anlayses results}

In the following tables, all predictors were categorical and dummy coded (reference level reported in parentheses). For post-hoc comparisons, Bonferroni-corrected \textit{p} values are reported.

\paragraph{Accuracy -- Only AI-support condition (all participants)}
\textcolor{white}{[parkour]}

\begin{table}[h]
\centering
\captionsetup{skip=5pt}
\begin{tabular}{lcccc}
\toprule

\multicolumn{5}{l}{\textbf{Regression coefficients}} \\
\midrule
Fixed effect
& Odds ratio
& 95\% CI
& \textit{z}
& \textit{p} \\
\midrule
Intercept
    & 3.93 & [2.49, 6.20] & 5.88 & < .001 \\
AI-support condition: LO (NS)
    & 1.42 & [0.95, 2.11] & 1.72 & .085 \\
AI-support condition: NIC (NS)
    & 1.85 & [1.24, 2.76] & 3.00 & .003 \\
AI-support condition: IC (NS)
    & 1.98 & [1.32, 2.96] & 3.33 & < .001 \\
\midrule

\addlinespace[1em]
\multicolumn{5}{l}{\textbf{ANOVA omnibus tests}} \\
\midrule
\multicolumn{2}{l}{Fixed effect}
& $\chi^2$
& df
& \textit{p} \\
\midrule
\multicolumn{2}{l}{AI-support condition}
    & 13.80 & 3 & .003 \\
\midrule

\addlinespace[1em]
\multicolumn{5}{l}{\textbf{Post-hoc contrasts}} \\
\midrule
Contrast
& Odds ratio
& \textit{SE}
& \textit{z}
& \textit{p} \\
\midrule
NS vs.\ LO
    & 0.71 & 0.14 & -1.72 & .513 \\
NS vs.\ NIC
    & 0.54 & 0.11 & -3.00 & .016 \\
NS vs.\ IC
    & 0.51 & 0.10 & -3.33 & .005 \\
LO vs.\ NIC
    & 0.77 & 0.16 & -1.28 & 1 \\
LO vs.\ IC
    & 0.72 & 0.15 & -1.62 & .637 \\
NIC vs.\ IC
    & 0.93 & 0.20 & -0.33 & 1 \\

\bottomrule
\end{tabular}
\caption{Results of the mixed-effects logistic regression on participants'
accuracy, including only AI-support condition as a fixed effect.}
\label{tab:study2-accuracy-simplemodel}
\end{table}

\newpage
\paragraph{Accuracy - Only AI-support condition (excluding non-interactive participants)}
\textcolor{white}{[parkour]}

\begin{table}[!h]
\centering
\captionsetup{skip=5pt}
\begin{tabular}{lcccc}
\toprule

\multicolumn{5}{l}{\textbf{Regression coefficients}} \\
\midrule
Fixed effect
& Odds ratio
& 95\% CI
& \textit{z}
& \textit{p} \\
\midrule
Intercept
    & 3.98 & [2.51, 6.32] & 5.87 & < .001 \\
AI-support condition: LO (NS)
    & 1.41 & [0.94, 2.12] & 1.68 & .092 \\
AI-support condition: NIC (NS)
    & 1.84 & [1.22, 2.76] & 2.93 & .003 \\
AI-support condition: IC (NS)
    & 2.69 & [1.68, 4.32] & 4.10 & < .001 \\
\midrule

\addlinespace[1em]
\multicolumn{5}{l}{\textbf{ANOVA omnibus tests}} \\
\midrule
\multicolumn{2}{l}{Fixed effect}
& $\chi^2$
& df
& \textit{p} \\
\midrule
\multicolumn{2}{l}{AI-support condition}
    & 18.96 & 3 & < .001 \\
\midrule

\addlinespace[1em]
\multicolumn{5}{l}{\textbf{Post-hoc contrasts}} \\
\midrule
Contrast
& Odds ratio
& \textit{SE}
& \textit{z}
& \textit{p} \\
\midrule
NS vs.\ LO
    & 0.71 & 0.15 & -1.68 & .555 \\
NS vs.\ NIC
    & 0.54 & 0.11 & -2.93 & .020 \\
NS vs.\ IC
    & 0.37 & 0.09 & -4.10 & < .001 \\
LO vs.\ NIC
    & 0.77 & 0.16 & -1.25 & 1 \\
LO vs.\ IC
    & 0.53 & 0.13 & -2.65 & .049 \\
NIC vs.\ IC
    & 0.68 & 0.17 & -1.56 & .710 \\

\bottomrule
\end{tabular}
\caption{Results of the mixed-effects logistic regression on participants'
accuracy, including only AI-support condition as a fixed effect.}
\label{tab:study2-accuracy-simplemodel-nononinteractive}
\end{table}

\newpage
\paragraph{Accuracy -- AI-support condition, item's ground truth, and model prediction accuracy (all participants)}
\textcolor{white}{[parkour]}
\begin{table}[!h]
\centering
\captionsetup{skip=5pt}
\begin{tabular}{lcccc}
\toprule

\multicolumn{5}{l}{\textbf{Regression coefficients}} \\
\midrule
Fixed effect
& Odds ratio
& 95\% CI
& \textit{z}
& \textit{p} \\
\midrule
Intercept
    & 1.58 & [0.69, 3.63] & 1.08 & .280 \\

AI-support condition: LO (NS)
    & 0.69 & [0.33, 1.45] & -0.99 & .324 \\

AI-support condition: NIC (NS)
    & 0.71 & [0.34, 1.49] & -0.90 & .368 \\

AI-support condition: IC (NS)
    & 1.20 & [0.57, 2.54] & 0.49 & .628 \\

Item ground truth (Le Conte's sparrow)
    & 3.33 & [1.03, 10.76] & 2.02 & .044 \\

Model prediction accuracy (correct)
    & 2.49 & [1.01, 6.14] & 1.98 & .048 \\

LO $\times$ Item ground truth
    & 0.43 & [0.16, 1.17] & -1.65 & .099 \\

NIC $\times$ Item ground truth
    & 0.36 & [0.13, 0.98] & -2.00 & .046 \\

IC $\times$ Item ground truth
    & 0.24 & [0.09, 0.65] & -2.81 & .005 \\

LO $\times$ Model prediction accuracy
    & 3.31 & [1.49, 7.35] & 2.94 & .003 \\

NIC $\times$ Model prediction accuracy
    & 6.20 & [2.71, 14.17] & 4.33 & < .001 \\

IC $\times$ Model prediction accuracy
    & 5.45 & [2.31, 12.83] & 3.88 & < .001 \\

Item ground truth $\times$ Model prediction accuracy
    & 0.30 & [0.08, 1.13] & -1.78 & .075 \\

LO $\times$ Item ground truth $\times$ Model prediction accuracy
    & 2.02 & [0.63, 6.43] & 1.18 & .236 \\

NIC $\times$ Item ground truth $\times$ Model prediction accuracy
    & 1.69 & [0.52, 5.52] & 0.87 & .382 \\

IC $\times$ Item ground truth $\times$ Model prediction accuracy
    & 1.22 & [0.37, 4.05] & 0.33 & .742 \\
\midrule

\addlinespace[1em]
\multicolumn{5}{l}{\textbf{ANOVA omnibus tests}} \\
\midrule
\multicolumn{2}{l}{Fixed effect}
& $\chi^2$
& df
& \textit{p} \\
\midrule
\multicolumn{2}{l}{Intercept}
    & 1.17 & 1 & .280 \\

\multicolumn{2}{l}{AI-support condition}
    & 3.09 & 3 & .377 \\

\multicolumn{2}{l}{Item ground truth}
    & 4.06 & 1 & .044 \\

\multicolumn{2}{l}{Model prediction accuracy}
    & 3.90 & 1 & .048 \\

\multicolumn{2}{l}{AI-support condition $\times$ Item ground truth}
    & 8.20 & 3 & .042 \\

\multicolumn{2}{l}{AI-support condition $\times$ Model prediction accuracy}
    & 23.73 & 3 & < .001 \\

\multicolumn{2}{l}{Item ground truth $\times$ Model prediction accuracy}
    & 3.17 & 1 & .075 \\

\multicolumn{2}{l}{AI-support condition $\times$ Item ground truth $\times$ Model prediction accuracy}
    & 1.69 & 3 & .640 \\
\midrule

\addlinespace[1em]
\multicolumn{5}{l}{\textbf{Post-hoc contrasts}} \\
\midrule
Contrast
& Estimate
& \textit{SE}
& \textit{z}
& \textit{p} \\
\midrule
\multicolumn{5}{l}{\textit{Model prediction wrong}} \\

NS vs.\ LO
    & 0.80 & 0.30 & 2.63 & .101 \\

NS vs.\ NIC
    & 0.85 & 0.30 & 2.82 & .058 \\

NS vs.\ IC
    & 0.54 & 0.30 & 1.77 & .919 \\

LO vs.\ NIC
    & 0.05 & 0.29 & 0.18 & 1 \\

LO vs.\ IC
    & -0.26 & 0.29 & -0.91 & 1 \\

NIC vs.\ IC
    & -0.31 & 0.29 & -1.09 & 1 \\

\addlinespace[0.5em]
\multicolumn{5}{l}{\textit{Model prediction correct}} \\

NS vs.\ LO
    & -0.75 & 0.22 & -3.37 & .009 \\

NS vs.\ NIC
    & -1.24 & 0.23 & -5.33 & < .001 \\

NS vs.\ IC
    & -1.26 & 0.24 & -5.34 & < .001 \\

LO vs.\ NIC
    & -0.49 & 0.24 & -2.04 & .498 \\

LO vs.\ IC
    & -0.51 & 0.24 & -2.10 & .429 \\

NIC vs.\ IC
    & -0.02 & 0.25 & -0.09 & 1 \\

\bottomrule
\end{tabular}
\caption{Results of the mixed-effects logistic regression on participants'
accuracy, including AI-support condition, item ground truth, model prediction
accuracy, and their interactions as fixed effects.}
\label{tab:study2-accuracy-fullmodel}
\end{table}

\newpage
\paragraph{Accuracy -- AI-support condition, item's ground truth, and model prediction accuracy (excluding non-interactive participants)}
\textcolor{white}{[parkour]}

\begin{table}[!h]
\centering
\captionsetup{skip=5pt}
\begin{tabular}{lcccc}
\toprule

\multicolumn{5}{l}{\textbf{Regression coefficients}} \\
\midrule
Fixed effect
& Odds ratio
& 95\% CI
& \textit{z}
& \textit{p} \\
\midrule
Intercept
    & 1.59 & [0.68, 3.74] & 1.06 & .288 \\

AI-support condition: LO (NS)
    & 0.69 & [0.32, 1.46] & -0.98 & .329 \\

AI-support condition: NIC (NS)
    & 0.71 & [0.34, 1.50] & -0.90 & .368 \\

AI-support condition: IC (NS)
    & 1.58 & [0.67, 3.74] & 1.05 & .295 \\

Item ground truth (Le Conte's)
    & 3.35 & [1.01, 11.17] & 1.97 & .049 \\

Model prediction accuracy (correct)
    & 2.50 & [0.99, 6.33] & 1.93 & .053 \\

LO $\times$ Item ground truth
    & 0.43 & [0.16, 1.16] & -1.66 & .096 \\

NIC $\times$ Item ground truth
    & 0.36 & [0.13, 0.98] & -2.01 & .045 \\

IC $\times$ Item ground truth
    & 0.23 & [0.07, 0.71] & -2.56 & .011 \\

LO $\times$ Model prediction accuracy
    & 3.32 & [1.49, 7.38] & 2.94 & .003 \\

NIC $\times$ Model prediction accuracy
    & 6.24 & [2.72, 14.28] & 4.33 & < .001 \\

IC $\times$ Model prediction accuracy
    & 6.01 & [2.13, 17.00] & 3.38 & < .001 \\

Item ground truth $\times$ Model prediction accuracy
    & 0.31 & [0.08, 1.17] & -1.73 & .084 \\

LO $\times$ Item ground truth $\times$ Model prediction accuracy
    & 2.03 & [0.63, 6.48] & 1.19 & .234 \\

NIC $\times$ Item ground truth $\times$ Model prediction accuracy
    & 1.70 & [0.52, 5.55] & 0.88 & .381 \\

IC $\times$ Item ground truth $\times$ Model prediction accuracy
    & 1.29 & [0.32, 5.27] & 0.35 & .724 \\
\midrule

\addlinespace[1em]
\multicolumn{5}{l}{\textbf{ANOVA omnibus tests}} \\
\midrule
\multicolumn{2}{l}{Fixed effect}
& $\chi^2$
& df
& \textit{p}\\
\midrule
\multicolumn{2}{l}{Intercept}
    & 1.13 & 1 & .288 \\

\multicolumn{2}{l}{AI-support condition}
    & 4.65 & 3 & .200 \\

\multicolumn{2}{l}{Item ground truth}
    & 3.88 & 1 & .049 \\

\multicolumn{2}{l}{Model prediction accuracy}
    & 3.73 & 1 & .053 \\

\multicolumn{2}{l}{AI-support condition $\times$ Item ground truth}
    & 7.28 & 3 & .064 \\

\multicolumn{2}{l}{AI-support condition $\times$ Model prediction accuracy}
    & 22.84 & 3 & < .001 \\

\multicolumn{2}{l}{Item ground truth $\times$ Model prediction accuracy}
    & 2.98 & 1 & .084 \\

\multicolumn{2}{l}{AI-support condition $\times$ Item ground truth $\times$ Model prediction accuracy}
    & 1.58 & 3 & .663 \\
\midrule

\addlinespace[1em]
\multicolumn{5}{l}{\textbf{Post-hoc contrasts}} \\
\midrule
Contrast
& Estimate
& \textit{SE}
& \textit{z}
& \textit{p} \\
\midrule
\multicolumn{5}{l}{\textit{Model prediction wrong}} \\
NS vs.\ LO
    & 0.80 & 0.31 & 2.62 & .106 \\

NS vs.\ NIC
    & 0.86 & 0.30 & 2.81 & .059 \\

NS vs.\ IC
    & 0.28 & 0.34 & 0.81 & 1 \\

LO vs.\ NIC
    & 0.05 & 0.29 & 0.19 & 1 \\

LO vs.\ IC
    & -0.52 & 0.33 & -1.57 & 1 \\

NIC vs.\ IC
    & -0.58 & 0.33 & -1.74 & .984 \\

\addlinespace[0.5em]
\multicolumn{5}{l}{\textit{Model prediction correct}} \\
NS vs.\ LO
    & -0.75 & 0.23 & -3.32 & .011 \\

NS vs.\ NIC
    & -1.24 & 0.24 & -5.26 & < .001 \\

NS vs.\ IC
    & -1.64 & 0.29 & -5.66 & < .001 \\

LO vs.\ NIC
    & -0.49 & 0.24 & -2.01 & .534 \\

LO vs.\ IC
    & -0.89 & 0.30 & -3.02 & .031 \\

NIC vs.\ IC
    & -0.40 & 0.30 & -1.33 & 1 \\

\bottomrule
\end{tabular}
\caption{Results of the mixed-effects logistic regression on participants'
accuracy, including AI-support condition, item ground truth, model prediction
accuracy, and their interactions as fixed effects.}
\label{tab:study2-accuracy-fullmodel-nononinteractive}
\end{table}

\newpage
\paragraph{Intervention behaviour -- Model concept-detection accuracy}
\textcolor{white}{[parkour]}

\begin{table}[!h]
\centering
\captionsetup{skip=5pt}
\begin{tabular}{lcccc}
\toprule

\multicolumn{5}{l}{\textbf{Regression coefficients}} \\
\midrule
Fixed effect
& Odds ratio
& 95\% CI
& \textit{z}
& \textit{p} \\
\midrule
Intercept
    & 0.17 & [0.09, 0.32] & -5.31 & < .001 \\

Model concept-detection accuracy: correct (wrong)
    & 0.06 & [0.04, 0.07] & -22.99 & < .001 \\

\bottomrule
\end{tabular}
\caption{Results of the mixed-effects logistic regression on participants' interaction behavior (i.e., whether they intervened at least once on a concept in a given trial), including model's concept detection accuracy as fixed effect.}
\label{tab:study2-intervention-modeldetection}
\end{table}

\begin{figure*}[!h]
\centering
\includegraphics[height=8.5cm]{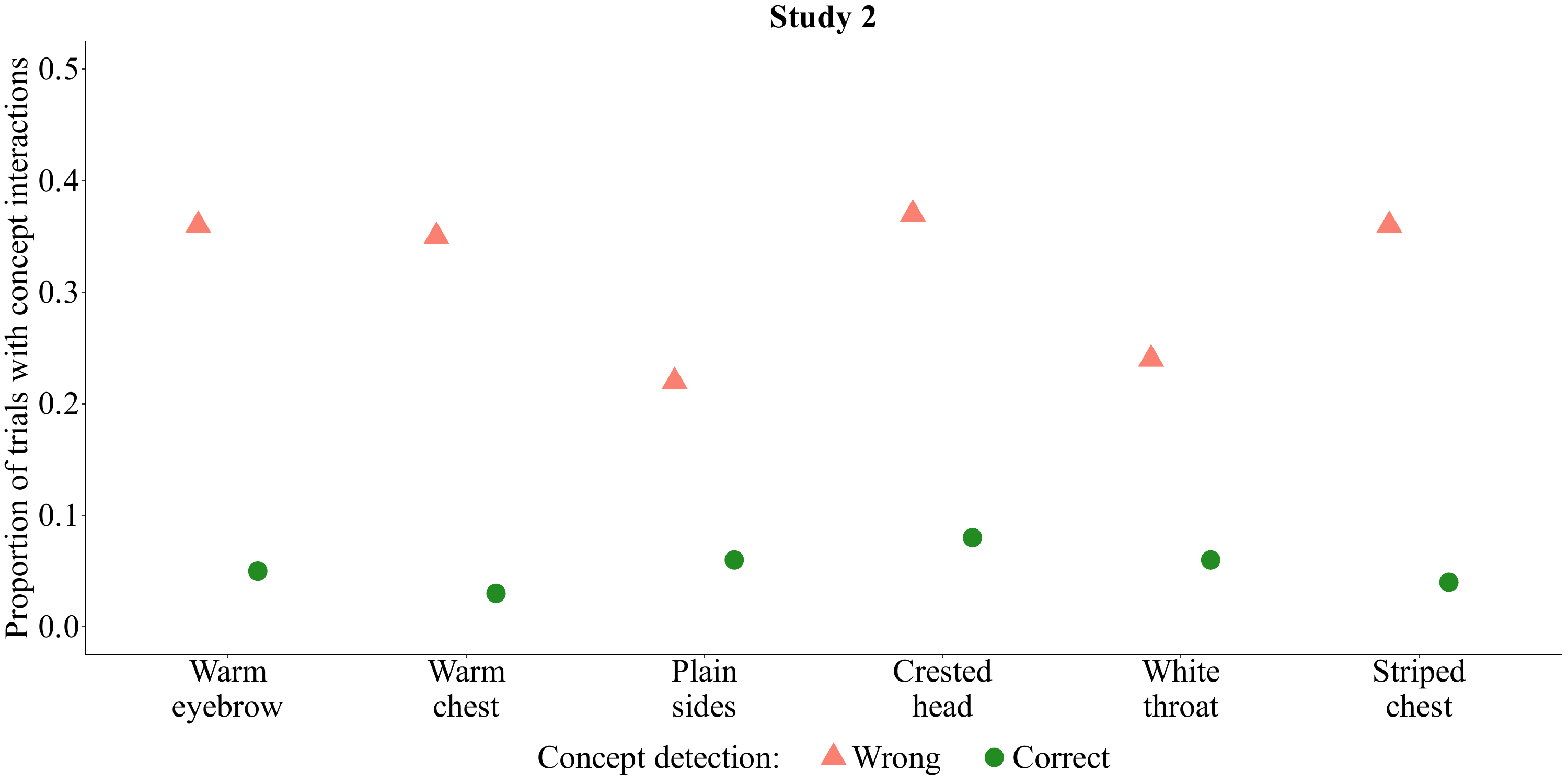}
\caption{Proportion of trials in which participants in the IC condition intervened on concept values, by model's concept detection accuracy (all participants included).}
\label{InteractionStudy2}
\end{figure*}

\paragraph{Intervention behavior - Model's concept detection accuracy (excluding non-interactive participants)}
\textcolor{white}{[parkour]}

\begin{table}[!h]
\centering
\captionsetup{skip=5pt}
\begin{tabular}{lcccc}
\toprule

\multicolumn{5}{l}{\textbf{Regression coefficients}} \\
\midrule
Fixed effect
& Odds ratio
& 95\% CI
& \textit{z}
& \textit{p} \\
\midrule
Intercept
    & 0.97 & [0.64, 1.46] & -0.16 & .874 \\

Model's concept detection accuracy: correct (wrong)
    & 0.06 & [0.05, 0.08] & -22.76 & < .001 \\

\bottomrule
\end{tabular}
\caption{Results of the mixed-effects logistic regression on participants' interaction behavior (i.e., whether they intervened at least once on a concept in a given trial), including model's concept detection accuracy as fixed effect.}
\label{tab:study2-intervention-modeldetection-nononinteractive}
\end{table}

\newpage
\paragraph{Intervention behavior - Impact of intervening on concept values on classification accuracy (all participants)}
\textcolor{white}{[parkour]}

\begin{table}[!h]
\centering
\captionsetup{skip=5pt}
\begin{tabular}{lcccc}
\toprule

\multicolumn{5}{l}{\textbf{Regression coefficients}} \\
\midrule
Fixed effect
& Odds ratio
& 95\% CI
& \textit{z}
& \textit{p} \\
\midrule
Intercept
    & 5.83 & [2.99, 11.40] & 5.16 & < .001 \\

Interaction within trial: true (false)
    & 2.78 & [1.73, 4.46] & 4.23 & < .001 \\

\bottomrule
\end{tabular}
\caption{Results of the mixed-effects logistic regression on participants' accuracy, including interaction behavior (i.e., having interacted at least once within a certain trial) as fixed effect.}
\label{tab:study2-intervention-accuracy}
\end{table}

\paragraph{Intervention behavior - Impact of intervening on concept values on classification accuracy (excluding non-interactive participants)}
\textcolor{white}{[parkour]}

\begin{table}[!h]
\centering
\captionsetup{skip=5pt}
\begin{tabular}{lcccc}
\toprule

\multicolumn{5}{l}{\textbf{Regression coefficients}} \\
\midrule
Fixed effect
& Odds ratio
& 95\% CI
& \textit{z}
& \textit{p} \\
\midrule
Intercept
    & 9.16 & [3.14, 26.69] & 4.06 & < .001 \\

Interaction within trial: true (false)
    & 2.89 & [1.46, 5.72] & 3.06 & .002 \\

\bottomrule
\end{tabular}
\caption{Results of the mixed-effects logistic regression on participants' accuracy, including interaction behavior (i.e., having interacted at least once within a certain trial) as fixed effect.}
\label{tab:study2-intervention-accuracy-nononinteractive}
\end{table}

\newpage
\paragraph{Confidence - AI-support condition (all participants)}
\textcolor{white}{[parkour]}

\begin{table}[!h]
\centering
\captionsetup{skip=5pt}
\begin{tabular}{lcccc}
\toprule

\multicolumn{5}{l}{\textbf{Regression coefficients}} \\
\midrule
Fixed effect
& \textit{b}
& \textit{SE}
& \textit{z}
& \textit{p} \\
\midrule
AI-support condition: LO (NS)
    & 0.77 & 0.33 & 2.29 & .022 \\

AI-support condition: NIC (NS)
    & 0.71 & 0.33 & 2.12 & .034 \\

AI-support condition: IC (NS)
    & 0.96 & 0.33 & 2.86 & .004 \\
\midrule

\addlinespace[1em]
\multicolumn{5}{l}{\textbf{ANOVA omnibus tests}} \\
\midrule
\multicolumn{2}{l}{Fixed effect}
& $\chi^2$
& df
& \textit{p} \\
\midrule
\multicolumn{2}{l}{AI-support condition}
    & 9.40 & 3 & .025 \\
\midrule

\addlinespace[1em]
\multicolumn{5}{l}{\textbf{Post-hoc contrasts}} \\
\midrule
Contrast
& Estimate
& \textit{SE}
& \textit{z}
& \textit{p} \\
\midrule
NS vs.\ LO
    & -0.77 & 0.33 & -2.29 & .134 \\

NS vs.\ NIC
    & -0.71 & 0.33 & -2.12 & .202 \\

NS vs.\ IC
    & -0.96 & 0.33 & -2.86 & .025 \\

LO vs.\ NIC
    & 0.06 & 0.33 & 0.17 & 1 \\

LO vs.\ IC
    & -0.19 & 0.33 & -0.58 & 1 \\

NIC vs.\ IC
    & -0.25 & 0.33 & -0.75 & 1 \\

\bottomrule
\end{tabular}
\caption{Results of the mixed-effects ordinal regression on participants' confidence ratings, including AI-support condition as fixed effect.}
\label{tab:study2-intervention}
\end{table}

\begin{figure*}[!h]
\centering
\includegraphics[height=7cm]{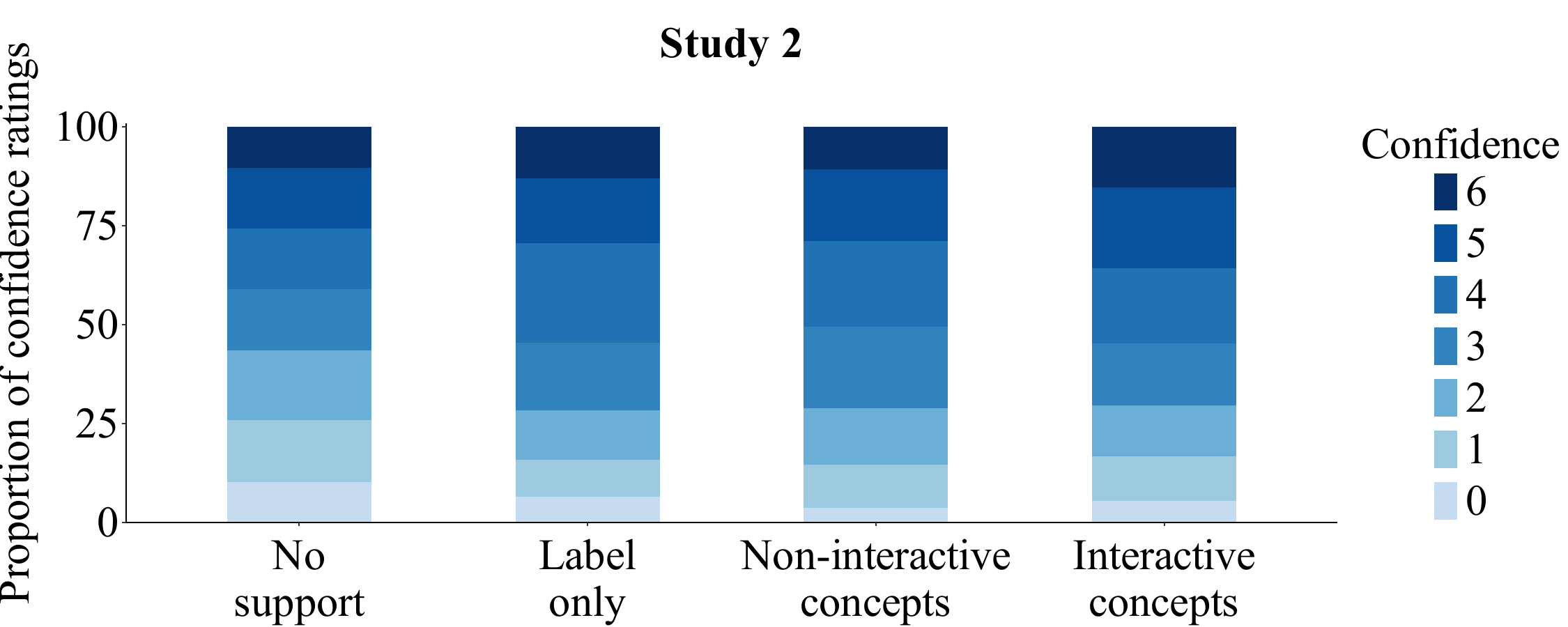}
\caption{Confidence ratings (0 = no confidence at all that the classification is correct; 6 = extreme confidence that the classification is correct) by AI-support condition (all participants included).}
\label{ConfidenceStudy2}
\end{figure*}

\newpage
\paragraph{Confidence -- AI-support condition (excluding non-interactive participants)}
\textcolor{white}{[parkour]}

\begin{table}[!h]
\centering
\captionsetup{skip=5pt}
\begin{tabular}{lcccc}
\toprule

\multicolumn{5}{l}{\textbf{Regression coefficients}} \\
\midrule
Fixed effect
& \textit{b}
& \textit{SE}
& \textit{z}
& \textit{p} \\
\midrule
AI-support condition: LO (NS)
    & 0.77 & 0.32 & 2.38 & .017 \\

AI-support condition: NIC (NS)
    & 0.71 & 0.32 & 2.21 & .027 \\

AI-support condition: IC (NS)
    & 1.07 & 0.36 & 2.96 & .003 \\
\midrule

\addlinespace[1em]
\multicolumn{5}{l}{\textbf{ANOVA omnibus tests}} \\
\midrule
\multicolumn{2}{l}{Fixed effect}
& $\chi^2$
& df
& \textit{p} \\
\midrule
\multicolumn{2}{l}{AI-support condition}
    & 10.50 & 3 & .015 \\
\midrule

\addlinespace[1em]
\multicolumn{5}{l}{\textbf{Post-hoc contrasts}} \\
\midrule
Contrast
& Estimate
& \textit{SE}
& \textit{z}
& \textit{p} \\
\midrule
NS vs.\ LO
    & -0.77 & 0.32 & -2.38 & .103 \\

NS vs.\ NIC
    & -0.71 & 0.32 & -2.21 & .164 \\

NS vs.\ IC
    & -1.07 & 0.36 & -2.96 & .019 \\

LO vs.\ NIC
    & 0.06 & 0.32 & 0.19 & 1 \\

LO vs.\ IC
    & -0.30 & 0.36 & -0.84 & 1 \\

NIC vs.\ IC
    & -0.36 & 0.36 & -1.00 & 1 \\

\bottomrule
\end{tabular}
\caption{Results of the mixed-effects ordinal regression on participants' confidence ratings, including AI-support condition as a fixed effect.}
\label{tab:study2-intervention-nononinteractive}
\end{table}

\newpage
\paragraph{Self-reported trust (all participants)}
\textcolor{white}{[parkour]}

\begin{table}[!h]
\centering
\captionsetup{skip=5pt}
\begin{tabular}{lccc}
\toprule

\multicolumn{4}{l}{\textbf{Kruskal--Wallis tests}} \\
\midrule
{}
& $\chi^2$
& df
& \textit{p} \\
\midrule
Trust index
    & 5.17 & 2 & .076 \\
Item 1
    & 1.07 & 2 & 1 \\
Item 2
    & 3.60 & 2 & 1 \\
Item 3
    & 5.32 & 2 & .559 \\
Item 4
    & 3.50 & 2 & 1 \\
Item 5
    & 4.48 & 2 & .853 \\
Item 6
    & 3.00 & 2 & 1 \\
Item 7
    & 9.85 & 2 & .058 \\
Item 8
    & 2.01 & 2 & 1 \\
\bottomrule
\end{tabular}
\caption{Results of the Kruskal-Wallis tests comparing the three AI-support conditions on the individual trust questionnaire items. The \textit{p} values were Bonferroni-adjusted for the eight items.}
\label{tab:study2-trust-items}
\end{table}

\paragraph{Self-reported trust (excluding non-interactive participants)}
\textcolor{white}{[parkour]}

\begin{table}[!h]
\centering
\captionsetup{skip=5pt}
\begin{tabular}{lccc}
\toprule

\multicolumn{4}{l}{\textbf{Kruskal--Wallis tests}} \\
\midrule
{}
& $\chi^2$
& df
& \textit{p} \\
\midrule
Trust index
    & 5.92 & 2 & .052 \\
Item 1
    & 1.12 & 2 & 1 \\
Item 2
    & 3.22 & 2 & 1 \\
Item 3
    & 6.46 & 2 & .316 \\
Item 4
    & 5.32 & 2 & .558 \\
Item 5
    & 2.32 & 2 & 1 \\
Item 6
    & 3.34 & 2 & 1 \\
Item 7
    & 10.22 & 2 & .048 \\
Item 8
    & 2.09 & 2 & 1 \\
\midrule

\addlinespace[1em]
\multicolumn{4}{l}{\textbf{Pairwise comparisons for Item 7}} \\
\midrule
Contrast
& {}
& {}
& \textit{p} \\
\midrule
LO vs.\ NIC
    & & & .528 \\
LO vs.\ IC
    & & & .005 \\
NIC vs.\ IC
    & & & .118 \\

\bottomrule
\end{tabular}
\caption{Results of the Kruskal--Wallis tests comparing the three AI-support conditions on the trust index and individual trust questionnaire items after excluding non-interactive participants. The \textit{p} values for the individual items were Bonferroni-adjusted for eight tests. Pairwise Wilcoxon rank-sum tests for Item 7 were Bonferroni-adjusted for three comparisons.}
\label{tab:study2-trust-items-nononinteractive}
\end{table}

\newpage
\section*{Computing infrastructure}
\addcontentsline{toc}{section}{Computing infrastructure}
\label{sec:infrastructure}

\bigskip
All experiments (CLIP/sentence-transformer encoding, SVM and logistic
regression training and evaluation) were run on a Linux
server:

\begin{table}[ht]
\centering
\captionsetup{skip=5pt}
\captionsetup{skip=5pt}
\begin{tabular}{ll}
\toprule
Component & Detail \\
\midrule
CPU & Intel Core i9-7920X @ 2.90GHz (12 cores / 24 threads) \\
RAM & 251\,GiB \\
GPU & 1$\times$ NVIDIA RTX A5000 (24\,GB) \\
\bottomrule
\end{tabular}
\caption{Hardware used for all reported experiments.}
\end{table}

\begin{table}[ht]
\centering
\begin{tabular}{ll}
\toprule
Library & Version \\
\midrule
Python & 3.12.13 \\
PyTorch & 2.5.1 (CUDA 12.1 build) \\
CLIP (\texttt{openai/CLIP}) & git commit \texttt{d05afc4} \\
sentence-transformers & 5.6.0 \\
scikit-learn & 1.9.0 \\
pandas & 3.0.3 \\
numpy & 2.5.1 \\
datasets (HuggingFace) & 5.0.0 \\
huggingface\_hub & 1.22.0 \\
langdetect & 1.0.9 \\
\bottomrule
\end{tabular}
\caption{Software versions}
\end{table}

Both SVM and
logistic-regression training run on CPU (scikit-learn), and CLIP/sentence-transformer encoding is the only step that uses the GPU.

\end{document}